\documentclass[aps,prx,10pt,twocolumn,superscriptaddress,nofootinbib,preprintnumbers,showpacs]{revtex4-2}
\usepackage{amsmath,amsfonts, amssymb, amsthm, dsfont}
\usepackage[T1]{fontenc}
\usepackage{yfonts}
\usepackage{bm}
\usepackage{bbm}
\usepackage{slashed}
\usepackage{mathrsfs}
\usepackage{graphicx}
\usepackage{verbatim}
\usepackage[bookmarks=true,colorlinks=true,linkcolor=blue, urlcolor=blue, citecolor=blue]{hyperref}
\hypersetup{linktocpage}
\usepackage{wasysym}
\usepackage[dvipsnames]{xcolor}
\usepackage{bbold}
\usepackage{braket}
\usepackage{diagbox} 
\usepackage{environ} 
\usepackage{relsize} 
\usepackage[caption=false]{subfig} 
\usepackage{tikz}
\usepackage{ulem}

\newcommand{\figref}[1]{Fig.~\ref{#1}}
\newcommand{\tabref}[1]{Tab.~\ref{#1}}

\newcommand{\beq}{\begin{equation}}
\newcommand{\eeq}{\end{equation}}
\newcommand{\beqn}{\begin{eqnarray}}
\newcommand{\eeqn}{\end{eqnarray}}
\DeclareMathAlphabet{\mathbbold}{U}{bbold}{m}{n}

\makeatletter
\newcommand\xleftrightarrow[2][]{%
\ext@arrow 9999{\longleftrightarrowfill@}{#1}{#2}}
\newcommand\longleftrightarrowfill@{%
\arrowfill@\leftarrow\relbar\rightarrow} \makeatother

\begin{document}

\title{Bipartite Fluctuations of Critical Fermi Surfaces}

\author{Xiao-Chuan Wu}
\affiliation{Kadanoff Center for Theoretical Physics, University of Chicago, Chicago, Illinois 60637, USA}

\begin{abstract}

Fluctuations of conserved quantities within a subsystem are non-local observables that provide unique insights into quantum many-body systems. In this paper, we study bipartite charge (and spin) fluctuations across interaction-driven ``metal-insulator transitions'' out of Landau Fermi liquids. The ``charge insulators'' include a class of non-Fermi-liquid states of fractionalized degrees of freedom, such as compressible composite Fermi liquids (for spinless electrons) and incompressible spin-liquid Mott insulators (for spin-$1/2$ electrons). We find that charge fluctuations $\mathcal{F}$ exhibit distinct leading-order scalings across the transition: $\mathcal{F} \sim L\log(L)$ in Landau Fermi liquids and $\mathcal{F} \sim L$ in charge insulators, where $L$ is the linear size of the subsystem. In composite Fermi liquids, under certain conditions, we also identify a universal constant term $-\textrm{f}(\theta)|\sigma_{xy}|/(2\pi)$ when the subsystem geometry contains a sharp corner, where $\textrm{f}(\theta)$ denotes a function of the corner angle, and $\sigma_{xy}$ is the Hall conductivity. At the critical point, provided the transition is continuous, the leading scaling $\mathcal{F}\sim L$ is accompanied by a subleading universal corner contribution $-\log(L)\textrm{f}(\theta)C_{\rho}/2$ with the same angle dependence $\textrm{f}(\theta)$, and the universal coefficient $C_{\rho}$ is directly related to the predicted universal jumps in longitudinal and Hall resistivities. These results establish fluctuation-transport relations, paving the way for numerical and experimental studies of unconventional quantum criticalities in metals.


\end{abstract}


\maketitle

\tableofcontents

\section{Introduction}

Measuring bipartite fluctuations of local observables provides a powerful approach to characterizing entanglement properties, symmetries, and correlations in various quantum phases and phase transitions. Although entanglement has become a fundamental organizing principle for the study of quantum matter~\cite{EECMP2016,WenRMP2017}, experimental measurement of many-body entanglement remains a challenge. Significant progress (see, e.g., Refs.~\cite{EEBF1,EEBF2,EEBF3,EEBF4,EEBF5,EEBF6,EEBF7,EEBF8,EEBF9,EEvsBF}) has been made toward establishing a direct relationship between entanglement entropy and the fluctuations of globally conserved quantities, such as particle number and magnetization. This relationship has been demonstrated to be feasible in special cases, such as one-dimensional Luttinger liquids~\cite{EEBF4,EEBF6} and free fermions in higher dimensions~\cite{EEBF7,EEBF6,EEBF2}. However, the connection between the two in strongly correlated systems beyond these examples is still poorly understood. The two quantities are observed to differ from each other in continuous symmetry-breaking phases, where the entanglement entropy still follows a boundary-law scaling while the bipartite fluctuations exhibit a multiplicative logarithmic enhancement to the boundary law.  Building on this observation, Ref.~\cite{BFQCP} proposed the use of the distinct scalings of bipartite fluctuations to detect conventional Landau phase transitions.

Despite decades of study, understanding unconventional quantum phases and phase transitions that are beyond Landau's symmetry-breaking paradigm remains a central problem in condensed matter physics. In recent years, conceptual breakthroughs~\cite{Nussinov2009SymTO,Gaiotto2015GeneSym} have been achieved by employing extended operators to define generalized symmetries and to characterize phases and phase transitions (see Refs.~\cite{gene_sym_review1,gene_sym_review2} and references therein). For example, abelian topological orders can be understood in terms of the condensation of 1-dimensional objects, leading to the spontaneous breaking of discrete $1$-form symmetries. Similarly, the Coulomb phase of electrodynamics can be interpreted as a spontaneous symmetry-breaking phase of $\textrm{U}(1)$ $1$-form symmetry, where gapless photons serve as Goldstone modes. This helps establish a generalized Landau symmetry paradigm. In this case, the criterion for determining whether the phase is symmetric or not is given by the scaling of disorder operators (or Wilson loops under duality). In this context, the concept of bipartite fluctuations is again closely related, which can be identified as the $\textrm{U}(1)$ disorder operator under the small-angle limit~\cite{wulog,chengdisop}. 

\begin{figure}
\includegraphics[width=0.32\textwidth]{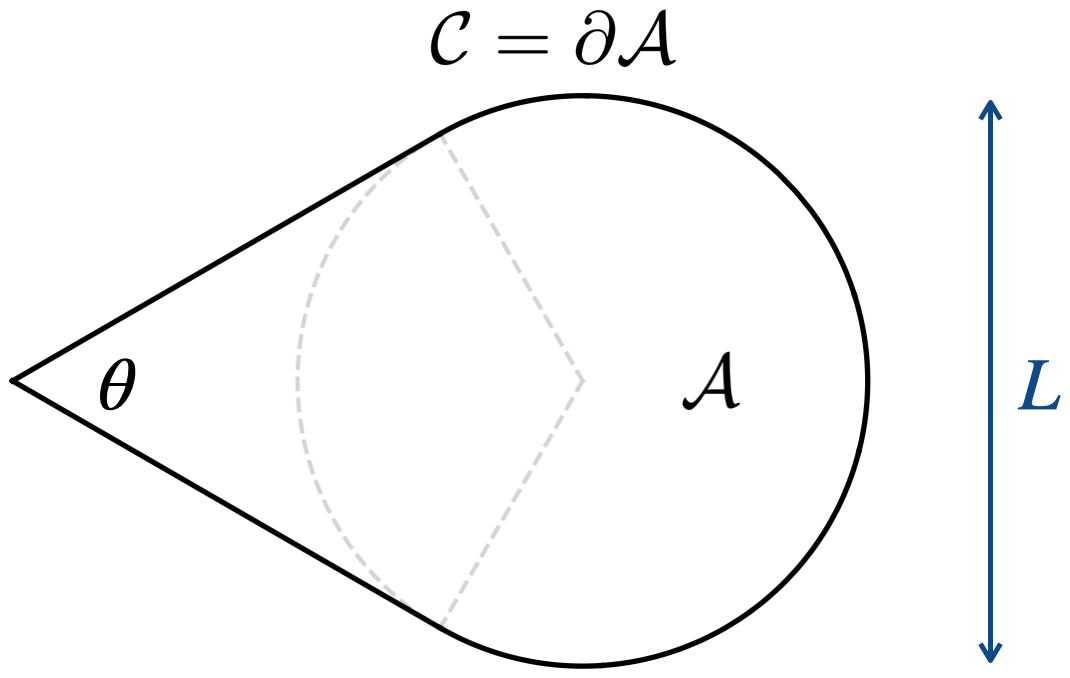}
\caption{Real-space subregion $\mathcal{A}$ with a single corner of opening angle $0<\theta<\pi$ and linear size $L$.} \label{fig:_corner}
\end{figure}

While extended operators play an important role in the conceptual understanding of phases of matter, recent studies have also shed light on intriguing quantitative aspects of disorder operators (and related bipartite fluctuations), particularly their scaling behaviors at conformally invariant quantum critical points~\cite{Dirac_log,wulog,chengdisop_Ising,chengdisop,uni_corner,Mengdisop1,Mengdisop2,Mengdisop3,Mengdisop4,Mengdisop5}. For instance, when considering the configuration of the subsystem as depicted in \figref{fig:_corner}, bipartite fluctuations exhibit a universal corner contribution with logarithmic scaling. This contribution has a universal angle dependence and is directly proportional to the current central charge $C_{J}$ of the conformal field theory (CFT)~\cite{Dirac_log,wulog,chengdisop,uni_corner}. Given the sometimes uncertain fate of proposed lattice realizations for exotic quantum critical points, numerical simulations, such as quantum Monte Carlo simulations, can assist in identifying whether these lattice models correspond to unitary CFTs by examining the sign of the universal corner contribution~\cite{Mengdisop1,Mengdisop2,Mengdisop3,Mengdisop4,Mengdisop5}. Moreover, Ref.~\cite{uni_corner} showed that the corner term of bipartite fluctuations in certain gapped systems, such as isotropic integer quantum Hall (QH) states and Laughlin states, is determined by the DC Hall conductivity.

Understanding quantum phase transitions in metals is significantly more challenging due to the abundance of low-energy excitations near the electronic Fermi surface. Even for conventional phase transitions associated with some form of broken symmetry in metals, the standard Hertz-Millis-Moriya framework~\cite{hertz1976,millis1993,HM_review} encounters serious difficulties in two spatial dimensions. Despite numerous attempts made in recent years~\cite{nayaknfl1,nayaknfl2,polchinskinfl,AIMnfl,nfl_response_1994,nfl2,nfl3,nfl4,nfl7,nfl8,nfl9,nfl10,nflUVIR,nflSYK1,nflSYK2,LU1_1,LU1_2,nlbosonization}, many aspects of the low-energy physics remain difficult to describe theoretically. A scenario that is both technically and conceptually more challenging is a continuous transition from an ordinary metal to an exotic gapless phase with abundant fractionalized excitations forming a Fermi surface. One such example is the proposed continuous Mott transition at half-filling~\cite{FlorensGeorges2004,lee2005,senthilmit1,senthilmit2} from a Landau Fermi liquid (FL) to a gapless Mott insulator (MI) with a spinon Fermi surface. This is potentially realized by the recent experimental observation~\cite{tmdmit1} of a continuous bandwidth-tuned transition from a metal to a paramagnetic Mott insulator in the transition metal dichalcogenide (TMD) moir{\'e} heterobilayer $\textrm{MoTe}_{2}/\textrm{WSe}_{2}$. However, the observed critical resistivity is anomalously large~\cite{tmdmit1}, exceeding the predictions of the original theory, at least in the clean limit\footnote{See Ref.~\cite{mitdis} for a discussion about the effects of disorder.}. To address this discrepancy, Ref.~\cite{frac_mit} has proposed a modified theory that explains the large critical resistivity by charge fractionalization.

Another notable example that has appeared in literature is the bandwidth-tuned transition from a Landau Fermi liquid (FL) to a composite Fermi liquid (CFL) describing the half-filled Landau level. The theoretical possibility of a continuous transition was originally proposed by Ref.~\cite{cflfl2} and later refined by Ref.~\cite{cflfl3} targeting moir{\'e} materials. The CFL phases at zero magnetic field were predicted to occur in twisted $\textrm{MoTe}_{2}$ at half and three-quarter fillings~\cite{tmdcfl1,tmdcfl2}.
On the experimental side, evidence for the CFL phase and the CFL-FL transition has been reported in both twisted $\textrm{MoTe}_{2}$ bilayers~\cite{CFLFL_Xu1,CFLFL_Xu2} and rhombohedral multilayer graphene~\cite{CFLFL_Ju}.

A general criterion~\cite{Kohn1964,SWM2000,Resta2002Rev,Resta2011Rev} distinguishing charge insulators from metals is the presence of a finite localization tensor, i.e., finite polarization fluctuations. Accordingly, as will be illustrated below, the CFL may be viewed as a ``charge insulator,'' and the CFL-FL transition can be interpreted as a ``metal-insulator transition'' in a broad sense. All of the above-mentioned ``metal-insulator transitions'' share a common theoretical framework, which we will review in Sec.~\ref{sec:_vrtx}, where electrons are fractionalized into fermionic and bosonic partons. The fermion sector forms a stable Fermi-surface state, while the boson sector contains a single relevant operator that drives the transition from a superfluid to a gapped state. Crucial insight into the possible transitions comes from the filling constraints of the bosons under translation symmetry and the Lieb-Schultz-Mattis theorem~\cite{frac_mit,cflfl3}. Without better terminology, we refer to this type of quantum critical points as {\it critical Fermi surfaces}, following Refs.~\cite{senthilmit1,senthilmit2}.

In this work, we point out the universal features of critical Fermi surfaces from the perspective of bipartite charge fluctuations $\mathcal{F}$. Although the transitions go beyond any symmetry principles, including the generalized ones~\cite{gene_sym_review1,gene_sym_review2}, the two phases can be distinguished by the distinct leading-order scalings $\mathcal{F}\sim L$ and $\mathcal{F}\sim L\log(L)$, just like conventional symmetry-breaking transitions~\cite{BFQCP}, where $L$ is the linear size of the subsystem under bipartition. Additionally, at the critical point, we find a subleading corner contribution with logarithmic scaling, reminiscent of the behavior seen in CFTs, despite the absence of conformal symmetry in such systems. The universal coefficient (denoted by $C_{\rho}$) of the logarithmic term can be directly linked to transport observables, such as the predicted longitudinal (and Hall) resistivity jump~\cite{senthilmit2,frac_mit,cflfl3} at the critical point. Furthermore, based on an isotropic field theory~\cite{HLR1993}, we predict a universal corner term in the non-Fermi-liquid phase of composite fermions governed by the DC Hall conductivity.

In contrast to conformally invariant quantum critical points, many powerful theoretical tools, such as the conformal bootstrap~\cite{bootstrapRMP,bootstrapRMP2}, are no longer applicable to strongly correlated metals. However, it is theoretically feasible in Monte Carlo simulations~\cite{Mengdisop1,Mengdisop2,Mengdisop3,Mengdisop4,Mengdisop5} to extract the subleading universal corner term of $\mathcal{F}$ using the method proposed in Ref.~\cite{Mengsubcor}. We anticipate that our findings, which establish a connection between the universal data $C_{\rho}$ of the critical points and charge fluctuations which are numerically and experimentally accessible, could aid future studies in identifying the existence and lattice realizations of these exotic quantum critical points in metals.

The remainder of the paper is organized as follows. In Sec.~\ref{sec:_BF_intro}, we provide relevant background and summarize our new findings. Sec.~\ref{sec:_vrtx} introduces a unified theoretical framework for the continuous phase transitions discussed in this work. In Sec.~\ref{sec:_CFLFL}, we explore spinless electrons, highlighting the intriguing behaviors of bipartite charge fluctuations within the CFL phase and at the CFL-FL transition. Sec.~\ref{sec:_Mott} shifts focus to spin-$1/2$ electrons, examining the universal behaviors of both charge and spin fluctuations across the Mott transitions. Finally, Sec.~\ref{sec:_Summary} offers further discussion of our results and suggests promising directions for future research.


\section{Bipartite Fluctuations}
\label{sec:_BF_intro}

In this section, we begin with a brief introduction to the preliminaries of bipartite fluctuations in various quantum states of matter, along with a general criterion for distinguishing charge insulators from metals. We then summarize our main findings on how bipartite fluctuations behave across a class of bandwidth-tuned continuous ``metal-insulator transitions.''

\subsection{Preliminaries I}

For any many-body system with a $\textrm{U}(1)$ global symmetry in $2+1$ dimensions, the bipartite fluctuations associated with a spatial subregion $\mathcal{A}$ are defined by\footnote{See App.~\ref{App:_Coulomb} for our conventions on the imaginary-time response function $\Pi^{\mu\nu}$ associated with the conserved Noether current $J^{\mu}$.}  
\begin{flalign}
\mathcal{F}_{\mathcal{A}}=\int_{\mathcal{A}}\textrm{d}^{2}\boldsymbol{x}\int_{\mathcal{A}}\textrm{d}^{2}\boldsymbol{y}\:\Pi^{\tau\tau}(\tau\rightarrow0,\boldsymbol{x}-\boldsymbol{y}),
\label{eq:_BF_def}
\end{flalign}
where $\Pi^{\tau\tau}$ denotes the equal-time density-density correlation function, i.e., the static structure factor.

In this introductory section, we focus on isotropic systems for simplicity and review several known results. More general and anisotropic systems have been discussed in Ref.~\cite{Dirac_log} and later in Refs.~\cite{Corner_Geom,Corner_Geom2}, where certain non-universal aspects are explored.


In App.~\ref{app:_BF_scaling}, we evaluate $\mathcal{F}_{\mathcal{A}}$ assuming a generic power-law instantaneous charge correlation in two dimensions
\begin{flalign}
\Pi^{\tau\tau}(\tau\rightarrow0,\boldsymbol{x})=\frac{-C_{0}}{|\boldsymbol{x}|^{\alpha}}.
\label{eq:_power-law_Pi00}
\end{flalign}
The resulting scaling behaviors are summarized in \tabref{tab:_BF_scaling}.

\begin{table}
\begin{tabular}{|c|c|c|}
\hline 
 & leading-order term & $\epsilon$-independent term \tabularnewline
\hline 
$2<\alpha<3$ & $L^{4-\alpha}$ & $L^{4-\alpha}$\tabularnewline
\hline 
$\alpha=3$ & $L\log (L)$ & $L\log (L)$\tabularnewline
\hline 
$3<\alpha<4$ & $L/\epsilon^{\alpha-3}$ & $L^{4-\alpha}$\tabularnewline
\hline 
$\alpha=4$ & $L/\epsilon$ & $\log (L)$
\tabularnewline
\hline 
$\alpha>4$ & $L/\epsilon^{\alpha-3}$ & $0$\tabularnewline
\hline 
\end{tabular}
\caption{
Scaling behaviors of the bipartite fluctuations $\mathcal{F}_{\mathcal{A}}$, based on the instantaneous density-density correlation Eq.~\eqref{eq:_power-law_Pi00}. Here, $L$ denotes the linear size of the subregion $\mathcal{A}$, and $\epsilon$ is a gauge-invariant UV cutoff introduced in the vortex theory (see App.~\ref{app:_BF_scaling}). Notable features include: the leading-order term follows boundary-law scaling for $\alpha > 3$; a universal subleading logarithmic term appears only when $\alpha = 4$ and the geometry of $\mathcal{A}$ includes sharp corners; and for $\alpha > 4$, the $\epsilon$-independent subleading term vanishes in the large-$L$ limit.}
\label{tab:_BF_scaling}
\end{table}

\begin{figure}
\includegraphics[width=0.24\textwidth]{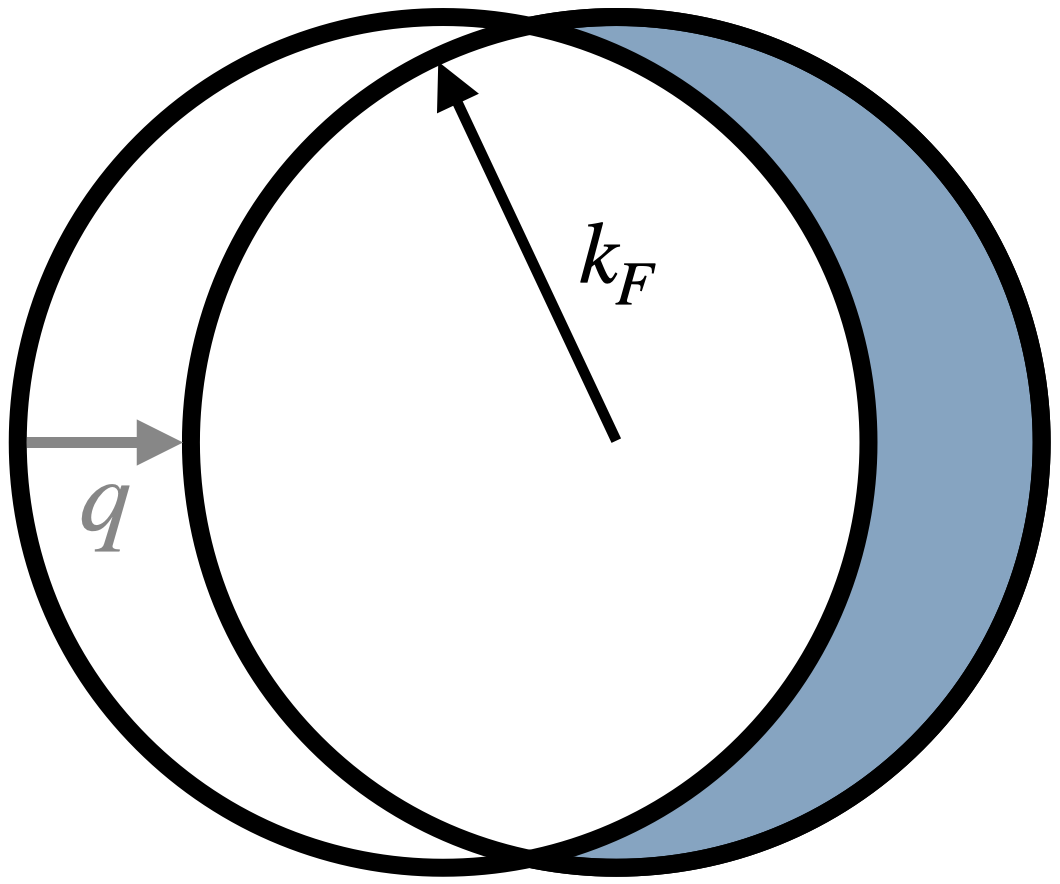}
\caption{Geometric interpretation of the static structure factor  $\Pi^{\tau\tau}(\tau\rightarrow0,\boldsymbol{q})$ for Fermi-surface states with LU(1) symmetry, where the shaded area indicates its value.} 
\label{fig:_ssf_geom}
\end{figure}

The simplest compressible state is the free Fermi gas. Using the free-fermion propagator, one can directly compute $\Pi^{\tau\tau}(\tau\rightarrow0,\boldsymbol{x})$ in real space and the static structure factor $\Pi^{\tau\tau}(\tau\rightarrow0,\boldsymbol{k})$ in momentum space
\begin{flalign}
\Pi^{\tau\tau}(\tau\rightarrow0,\boldsymbol{x})&=\frac{k_{F}}{4\pi^{3}}\frac{-1}{|\boldsymbol{x}|^{3}},\nonumber\\\Pi^{\tau\tau}(\tau\rightarrow0,\boldsymbol{k})&=\frac{k_{F}|\boldsymbol{k}|}{2\pi^{2}},
\label{eq:_FS_Pi00}
\end{flalign}
where $k_{F}$ denotes the Fermi momentum. According to \tabref{tab:_BF_scaling}, we observe the scaling $\mathcal{F}\sim L\log(L)$. This is a well-known result closely related to the entanglement entropy of free fermions~\cite{FSEE1,FSEE2,FSEE_Yang,FSEE_swingle1,EEBF7,EEBF6,EEBF2,FSEE_Ryu}. Recent studies~\cite{Tam2022, Tam2024} have further explored related calculations addressing the role of Fermi-surface topology in free-fermion systems. In App.~\ref{App:_LU1}, we reproduce Eq.~\eqref{eq:_FS_Pi00} based the $\textrm{LU}(1)$ anomaly of Fermi surface states~\cite{LU1_1}, which remains valid even in the presence of strong forward scattering that preserves $\textrm{LU}(1)$ symmetry. This approach also provides a geometric interpretation~\cite{Prashant} of the static structure factor in terms of the area difference induced by shifting the Fermi surface, as illustrated in Fig.~\ref{fig:_ssf_geom}.

Another simple compressible state is provided by the superfluid phase. Let $\phi$ denote the $\mathrm{U}(1)$ order parameter. The gapless Goldstone mode $\theta$ describes the phase fluctuations of $\phi=\rho_{s}e^{\mathtt{i}\theta}$, where $\rho_{s}$ is the superfluid stiffness. Power counting of the action $\int\textrm{d}^{3}x(\partial\theta)^{2}$ yields a scaling dimension of $\Delta[\partial\theta]=3/2$ for the $\textrm{U}(1)$ charge density, implying that the equal-time density-density correlation decays as $|\boldsymbol{x}|^{-3}$. Once again, one finds the scaling of bipartite fluctuations $\mathcal{F}\sim L\log(L)$.

An important class of incompressible states is given by unitary CFTs. The two-point function of the conserved spin-$1$ current has a rigid structure~\cite{bootstrapRMP}
\begin{flalign}
\left\langle J^{\mu}(x)J^{\nu}(0)\right\rangle =\frac{C_{J}}{|x|^{4}}\left(\delta^{\mu\nu}-\frac{2x^{\mu}x^{\nu}}{|x|^{2}}\right), \label{eq:_JJ_CFT}
\end{flalign}
where the scaling dimension of $J^{\mu}$ is protected. The overall coefficient $C_{J}>0$, known as the current central charge, is a universal data of the CFT and is related to the longitudinal conductivity via $\sigma_{xx}=\frac{\pi^{2}}{2}C_{J}$. Considering the subsystem with a single corner, as shown in \figref{fig:_corner}, one can verify the scaling behavior~\cite{wulog,uni_corner,chengdisop} 
\begin{flalign}
\mathcal{F}=\#L-\frac{C_{J}}{2}\textrm{f}(\theta)\log(L)+\ldots.
\label{eq:_BF_CFT}
\end{flalign}
where $\#$ is a non-universal number depending on the UV cut-off, and the universal angle dependence
\begin{equation} 
\textrm{f}(\theta)=1+(\pi-\theta)\cot(\theta).
\label{eq:_angle_function}
\end{equation}
applies to any 2+1D CFTs. This corner contribution has also been previously discussed in the context of non-interacting Dirac systems~\cite{Dirac_log}. However, the question remains whether the universal behavior described in Eq.~\eqref{eq:_BF_CFT} could also exist at certain quantum critical points lacking conformal symmetry.

Another standard class of incompressible states is insulators with a charge gap. In these systems, the equal-time density-density correlation decays exponentially, leading to a boundary-law scaling of bipartite fluctuations. An intriguing observation from Ref.~\cite{uni_corner} is that certain gapped QH insulators with continuous translational and rotational symmetries exhibit a universal corner contribution when the subregion geometry includes a sharp corner, as in \figref{fig:_corner}:  
\begin{equation} 
\mathcal{F}=\#L-\frac{|\sigma_{xy}|}{2\pi}\textrm{f}(\theta)+\ldots.
\label{eq:_BF_QH}
\end{equation}
The angle dependence is once again given by Eq.~\eqref{eq:_angle_function}, and $\sigma_{xy}=\nu/(2\pi)$ represents the DC Hall conductivity, where $\nu$ denotes the filling factor. For integer fillings $\nu\in\mathbb{Z}$, the corner term can be derived analytically, while for the fractional QH insulator at $\nu=1/3$, it has been verified using Monte Carlo simulations based on the Laughlin wavefunction~\cite{uni_corner}. It would be interesting to explore whether a similar corner term exists in compressible QH states at even-denominator fillings and to understand the conditions under which the formula in Eq.~\eqref{eq:_BF_QH} holds.

\subsection{Preliminaries II}
\label{subsec:_SSF_SWM}

\begin{table}
\begin{tabular}{|c|c|c|c|}
\hline 
 & FL & MI & CFL\tabularnewline
\hline 
Compressibility & finite & zero & finite\tabularnewline
\hline 
Drude weight  & finite & zero & zero\tabularnewline
\hline 
Localization tensor  & divergent  & finite & finite\tabularnewline
\hline 
Charge fluctuations & $\mathcal{F}^{c}\sim L\log(L)$ & $\mathcal{F}^{c}\sim L$ & $\mathcal{F}^{c}\sim L$\tabularnewline
\hline 
\end{tabular}

\caption{Distinct behaviors of various physical quantities across the transition~\cite{senthilmit1,senthilmit2,frac_mit} from a Landau Fermi liquid (FL) to a Mott insulator (MI), and the transition~\cite{cflfl2,cflfl3,cfl_drude} from a Landau FL to a composite Fermi liquid (CFL). This work highlights the different behaviors of the localization tensor (i.e., polarization fluctuations) and the leading-order scaling of charge fluctuations.}
\label{tab:_FL_MI_CFL}

\end{table}

In this subsection, we clarify some terminology that will be useful in presenting our results. By definition, bipartite fluctuations depend on the inversion-symmetric part of the static structure factor
\begin{equation} 
\Pi_{+}^{\tau\tau}(\tau\rightarrow0,\boldsymbol{k})=\frac{\Pi^{\tau\tau}(\tau\rightarrow0,\boldsymbol{k})+\Pi^{\tau\tau}(\tau\rightarrow0,-\boldsymbol{k})}{2}.
\label{eq:_SSF}
\end{equation}
Its long-wavelength expansion can be expressed as
\begin{equation} 
\Pi_{+}^{\tau\tau}(\tau\rightarrow0,\boldsymbol{k})=\mathsf{G}^{ij}k_{i}k_{j}+\mathscr{O}(|\boldsymbol{k}|^{3}),
\label{eq:_SSF_expd}
\end{equation}
where $\mathsf{G}^{ij}$ represents the real part of the localization tensor (i.e., polarization fluctuations)~\cite{Kohn1964,SWM2000,Resta2002Rev,Resta2011Rev}. Recently, $\mathsf{G}^{ij}$ has also been referred to as the ``quantum weight'' in Refs. \cite{Fu2024-1,Fu2024-2,Fu2024-3}. In this work, we are interested in phase transitions from a Landau FL, where $\mathsf{G}^{ij}$ is divergent, to another phase where $\mathsf{G}^{ij}$ becomes finite. According to the general criterion in Ref.~\cite{Kohn1964,SWM2000,Resta2002Rev,Resta2011Rev,Hetenyi2022}, systems with a finite value of $\mathsf{G}^{ij}$ are classified as charge insulators. 


Further physical insights into the localization tensor $\mathsf{G}^{ij}$ can be gained by considering the Souza-Wilkens-Martin (SWM) sum rule~\cite{SWM2000}
\begin{flalign}
\mathsf{G}^{ij}=\int_{0}^{+\infty}\frac{\textrm{d}\omega}{\pi}\frac{\textrm{Re}\sigma_{+}^{ij}(\omega)}{\omega},
\label{eq:_SWM_sum}
\end{flalign}
where $\sigma_{+}^{ij}=(\sigma^{ij}+\sigma^{ji})/2$ denotes the longitudinal conductivity. It establishes a fluctuation-dissipation theorem that connects the ground-state fluctuations of polarization to the optical conductivity. The linear behavior $\Pi^{\tau\tau}(\tau\rightarrow0,\boldsymbol{k})\sim|\boldsymbol{k}|$ observed in Landau FLs (see also Eq.~\eqref{eq:_FS_Pi00} for the free Fermi gas) can be explained by the presence of a nonzero Drude weight $D^{ij}$ such that $\sigma_{+}^{ab}(\omega)\supset D^{ij}(\delta(\omega)+\frac{\mathtt{i}}{\pi\omega})$, leading to a linear divergence in $\mathsf{G}^{ij}\sim1/|\boldsymbol{k}|$ due to the identical scaling of frequency $\omega$ and
momentum $|\boldsymbol{k}|$. It is worth noting that a vanishing Drude weight does not necessarily imply a finite value of $\mathsf{G}^{ij}$. An example of this is provided by CFTs, which are incompressible, where the localization tensor diverges logarithmically as $\mathsf{G}^{ij} \sim \log(1/|\boldsymbol{k}|)$.

In \tabref{tab:_FL_MI_CFL}, we summarize the behaviors of several physical quantities across the ``metal-insulator transitions'' where the ``charge insulators'' include both compressible and incompressible examples. Given that the primary focus of this work is on the universal field theories describing phase transitions, we assume continuous rotational symmetry for simplicity. This assumption renders the localization tensor isotropic, such that $\mathsf{G}^{ij}=\mathsf{G}\delta^{ij}$, where
\begin{flalign}
\mathsf{G}=\frac{\textrm{Tr}(\mathsf{G}^{ij})}{2}=\sqrt{\det(\mathsf{G}^{ij})}.
\end{flalign}
Charge insulators with a general anisotropic $\mathsf{G}^{ij}$ will be addressed in an upcoming work~\cite{Corner_Geom}.

\subsection{Summary of Results}


If the above-mentioned phase transitions are indeed continuous, the critical points are expected to be incompressible, with a vanishing Drude weight and intriguing universal behaviors in the localization tensor and charge fluctuations, which will be summarized below. The descriptions of these continuous phase transitions are unified within a theoretical framework that introduces vortex excitations in Landau FLs, as will be discussed in Sec.~\ref{sec:_vrtx}. Different charge insulators can then be realized by placing vortices in different quantum states. 



Below, we summarize the key results in Sec.~\ref{sec:_CFLFL} and Sec.~\ref{sec:_Mott}, where we consider examples of clean electron systems at half-filling, assuming disorder effects are weak.

\begin{figure}
\includegraphics[width=0.42\textwidth]{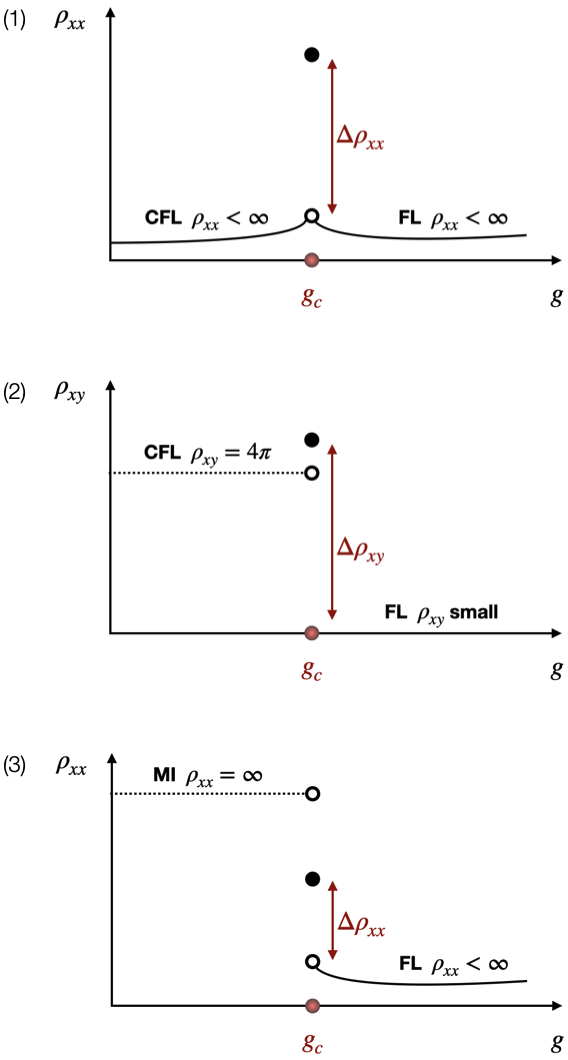}
\caption{Panels (1) and (2) show the predicted universal jumps in longitudinal resistivity $\Delta\rho_{xx}$
and Hall resistivity $\Delta\rho_{xy}$ at the transition from a Fermi liquid (FL) to a composite Fermi liquid (CFL)~\cite{cflfl3}; panel (3) illustrates the predicted universal longitudinal resistivity jump $\Delta\rho_{xx}$ at the transition from a Fermi liquid (FL) to a Mott insulator (MI)~\cite{senthilmit2,frac_mit}. In all cases, the tuning parameter $g$ corresponds to the electron bandwidth. These resistivity jumps are directly connected to the universal corner contribution to bipartite charge fluctuations, as given in Eq.~\eqref{eq:_C_jump}.} 
\label{fig:_jump}
\end{figure}

\begin{enumerate}
    \item 
Despite being compressible, CFLs exhibit the leading-order boundary-law scaling $\mathcal{F}\sim L$ for bipartite charge fluctuations, in contrast to the $\mathcal{F}\sim L\log(L)$ scaling seen in other compressible phases like Landau FLs and superfluids. This difference arises because the instantaneous charge correlation from the gapless modes decays as $|\boldsymbol{x}|^{-5}$ (up to logarithmic corrections). In addition, the gapped modes give rise to a subleading corner term analogous to those found in gapped QH insulators~\cite{uni_corner}. For the subsystem geometry depicted in \figref{fig:_corner}, the final result takes the form
\begin{flalign}
\mathcal{F}=\#L-\frac{\mathsf{G}}{\pi}\textrm{f}(\theta)+\ldots
\label{eq:_BF_QH_2}
\end{flalign}
where $\#$ represents a non-universal constant, and $\mathsf{G}^{ij}=\mathsf{G}\delta^{ij}$ is the localization tensor introduced in Eq.~\eqref{eq:_SSF_expd} and Eq.~\eqref{eq:_SWM_sum}. Based on the Halperin-Lee-Read (HLR) theory~\cite{HLR1993}, we find the coefficient 
\begin{flalign}
\mathsf{G}=\frac{|\sigma_{xy}|}{2}
\label{eq:_QM_Hall}
\end{flalign}
where $\sigma_{xy}=1/(4\pi)$
is the DC Hall conductivity of the half-filled Landau level. The angle dependence remains governed by the ``super-universal'' formula Eq.~\eqref{eq:_angle_function}. Note that, in general, certain conditions must be satisfied for Eq.~\eqref{eq:_QM_Hall} to hold in charge insulators~\cite{Corner_Geom}, as will be mentioned in Sec.~\ref{subsubsec:_BF_CFL}.

    \item
At the critical point of the continuous CFL-FL transition, we find that the long-wavelength behavior of the static structure factor follows a CFT-like scaling, leading to a universal divergence of the localization tensor
\begin{flalign}
\mathsf{G}=\frac{\pi C_{\rho}}{2}\log(\xi)+\ldots
\label{eq:_QM_log-div}
\end{flalign}
where $\xi$ is the correlation length of the system. The coefficient $C_{\rho}$ should be understood as universal data associated with the critical Fermi surface\footnote{We use a different notation $C_{\rho}$ for critical Fermi surfaces to distinguish it from $C_{J}$ used in CFTs.}. An important phenomenology of this type of transition is the universal jumps in longitudinal resistivity $\Delta\rho_{xx}(\omega/T)$ and Hall resistivity $\Delta\rho_{xy}(\omega/T)$ at the critical point~\cite{cflfl3}, where $\omega$ is frequency and $T$ is temperature, as depicted schematically in \figref{fig:_jump}. In the limit $\omega/T\rightarrow\infty$, these jumps are related to the universal coefficient $C_{\rho}$ through 
\begin{flalign}
C_{\rho}=\frac{2}{\pi^{2}}\frac{\Delta\rho_{xx}}{(\Delta\rho_{xx})^{2}+(\Delta\rho_{xy})^{2}}.
\label{eq:_C_jump}
\end{flalign}
Despite the absence of conformal symmetry in the system, the universal behavior in Eq.~\eqref{eq:_QM_log-div} results in the scaling of bipartite fluctuations similar to that seen in CFTs (see Eq.~\eqref{eq:_BF_CFT})
\begin{flalign}
\mathcal{F}=\#L-\frac{C_{\rho}}{2}\textrm{f}(\theta)\log(L)+\ldots.
\label{eq:_BF_CFS}
\end{flalign}

    \item 
The continuous Mott transition~\cite{senthilmit2} can be detected through the distinct leading-order scalings of bipartite charge fluctuations: $\mathcal{F}^{c} \sim L\log(L)$ in the metallic phase and $\mathcal{F}^{c} \sim L$ in the insulating phase. At the non-conformal critical point, the localization tensor exhibits a universal divergence, as described in Eq.~\eqref{eq:_QM_log-div}. The scaling of bipartite charge fluctuations $\mathcal{F}^{c}$ follows Eq.~\eqref{eq:_BF_CFS} with the coefficient $C_{\rho}$ given by the current central charge $C_{J}^{(\textrm{XY})}$ of the 3D XY universality class. Since $\Delta\rho_{xy}=0$, the resistivity jump is entirely longitudinal, and Eq.~\eqref{eq:_C_jump} remains applicable. The system also possesses a spin $\textrm{U}(1)$ symmetry, allowing for the definition of bipartite spin fluctuations $\mathcal{F}^{s}$. In both phases, the Fermi-surface states of spin excitations manifest as $\mathcal{F}^{s}\sim L\log(L)$. Richer physics can be revealed by extending the vortex theory in Sec.~\ref{sec:_vrtx} to include both charge and spin vortices. Further insights into the associated multicritical behavior will be discussed in Sec.~\ref{subsec:_Mott_sc}.

    \item 
In the modified proposal~\cite{frac_mit} of the continuous Mott transition involving charge fractionalization, the leading-order scaling of bipartite charge fluctuations $\mathcal{F}^c$ remains the same: $\mathcal{F}^c \sim L \log L$ in the metallic phase and $\mathcal{F}^c \sim L$ in the insulating phase. However, at the quantum critical point, the universal corner contribution in Eq.~\eqref{eq:_BF_CFS} is suppressed relative to the original theory~\cite{senthilmit2}, with $C_{\rho}=2C_{J}^{(\textrm{XY})}/N^{2}$, where $N$ is an even integer corresponding to the fractional electric charge $e/N$ carried by each charge carrier. Consequently, in the time-reversal-invariant version of Eq.~\eqref{eq:_C_jump} with $\Delta\rho_{xy}=0$, the critical longitudinal resistivity jump $\Delta\rho_{xx}$ is enhanced by a factor of $N^{2}/2$, consistent with the large critical resistivity observed experimentally~\cite{tmdmit1}. Another key difference from the original theory is that the bipartite spin fluctuations $\mathcal{F}^{s}$ obey boundary-law scaling $\mathcal{F}^{s} \sim L$ in the Mott insulating phase.

\end{enumerate}

\section{Vortex Theory Framework} \label{sec:_vrtx}

In this section, we present a general theoretical framework for describing a class of continuous quantum phase transitions out of Landau FLs, which will be utilized repeatedly in this paper.

One approach to quantum phase transitions is to incorporate vortex excitations into the Landau FL phase of electrons. If the vortex sector is not trivially gapped, the system is driven into a different phase. To formalize this idea, it is often convenient to begin with a parton decomposition of the spin-$1/2$ or spinless electron operator $c({\boldsymbol{x}})$, expressed as
\begin{flalign}
c({\boldsymbol{x}})=b({\boldsymbol{x}})f({\boldsymbol{x})}.
\label{eq:_parton_1}
\end{flalign}
where the assignment of the spin quantum number to the fermionic and bosonic partons $f$ and $b$ varies depending on specific examples, as summarized in \tabref{tab:_parton}. Notably, as will be discussed in Sec.~\ref{sec:_Summary}, the same physics can be realized without introducing partons, by using nonlinear bosonization~\cite{nlbosonization}.

For illustrative purposes, let us focus on the case of spinless electrons. The generalization to spinful cases follows in a similar way. Due to the $\textrm{U}(1)$ gauge redundancy in Eq.~\eqref{eq:_parton_1}, both $b$ and $f$ are coupled to a dynamical $\textrm{U}(1)$ gauge field $a_{\mu}$, carrying equal and opposite charges of $\pm1$. When a background field $A_{\mu}$ is introduced for the global $\textrm{U}(1)$ symmetry, the effective Lagrangian can be schematically expressed as
\begin{flalign}
\mathcal{L}=\mathcal{L}[f,a+e_{f}A]+\mathcal{L}[b,-a+e_{b}A]+\ldots\label{eq:_parton_theory}
\end{flalign}
where $e_{f} + e_{b} = 1$ is the $\textrm{U}(1)$ charge carried by the electron. As demonstrated in App.~\ref{App:_Ioffe-Larkin}, the gauge-invariant response remains independent of the specific charge assignment $(e_{f},e_{b})$. For simplicity, we can choose $(e_{f}, e_{b}) = (0, 1)$ and refer to the bosonic parton $b$ as the chargon.

The next step is to go to the dual vortex representation of the chargon sector 
\begin{flalign}
\mathcal{L}=\mathcal{L}_{\textrm{FS}}[f,a]+\frac{\mathtt{i}}{2\pi}\tilde{a}\wedge\textrm{d}(A-a)+\mathcal{L}_{\textrm{vrtx}}[v,\tilde{a}]+\ldots
\label{eq:_parton_vrtx}
\end{flalign}
where the flux of the gauge field $\tilde{a}$ represents the density of $b$. It is important to note that under gauge constraint, the density of $b$ equals to the density of $f$, as well as the density of electrons. We are interested in the parton mean-field state (before turning on gauge-field fluctuations), where the $f$-fermions occupy the same Fermi-surface state as the original electrons. The Landau FL phase of electrons can be reproduced when the vortices $v$ are trivially gapped. In this case, the Maxwell term $\frac{1}{2e^{2}}\textrm{d}\tilde{a}\wedge\star\textrm{d}\tilde{a}$ becomes important, and the equation of motion of $\tilde{a}$ leads to a mass term $\frac{e^{2}}{8\pi^{2}}(a_{\mu})^{2}$ for the gauge field $a_{\mu}$. In the IR, the Fermi-surface state of $f$-fermions essentially corresponds to the ordinary metallic phase of gauge-invariant electrons. Conversely, various interesting electronic phases can be realized by placing the vortices $v$ in different states. For instance, incompressible Mott insulators~\cite{senthilmit1,senthilmit2,frac_mit} can be realized when the vortices are in the Higgs phase or certain topological orders. Compressible CFLs~\cite{cflfl1,cflfl2,cflfl3} can also be realized by putting the vortices in integer QH states. Although we do not consider translation symmetry breaking in this paper, density-wave states~\cite{senthilmitdw, frac_mit, cflfl3} can also be conveniently described by the theory in Eq.~\eqref{eq:_parton_vrtx}, where the vortex band structure has multiple minima in the Brillouin zone, and vortex condensation results in lattice translation symmetry breaking.

\begin{widetext}

\begin{table}
\begin{tabular}{|c|c|c|c|c|}
\hline 
 & {\it(1)} FL-CFL transition~\cite{cflfl2,cflfl3} & {\it(2)} Mott transition~\cite{senthilmit1,senthilmit2} & {\it(3)} Mott transition~\cite{frac_mit} & $\ldots$\tabularnewline
\hline 
electron $c$ & spinless $c(\boldsymbol{x})=b(\boldsymbol{x})f(\boldsymbol{x})$ & spinful $c_{\sigma}(\boldsymbol{x})=b(\boldsymbol{x})f_{\sigma}(\boldsymbol{x})$ & spinful $c_{\sigma}(\boldsymbol{x})=b_{\sigma}(\boldsymbol{x})f_{\sigma}(\boldsymbol{x})$ & $\ldots$\tabularnewline
\hline 
fermion $f$ & Fermi surface & Fermi surface & Fermi surface & Fermi surface\tabularnewline
\hline 
boson $b$ & superfluid to Laughlin at $\nu=1/2$ & superfluid to trivial gapped & superfluid to $Z_{N}$ topological order & superfluid to ...\tabularnewline
\hline 
vortex $v$ & trivial gapped to IQH at $\nu=-2$ & trivial gapped to Higgs & trivial gapped to Higgs $\textrm{U}(1)\rightarrow Z_{N}$ & trivial to ...\tabularnewline
\hline 
\end{tabular}
\caption{Summary of the parton constructions for continuous quantum phase transitions out of spinless and spin-$1/2$ Landau Fermi liquids at half-filling. Here, $\sigma=\uparrow,\downarrow$ denotes the spin quantum number. The dual vortex field $v$ is introduced in Eq.~\eqref{eq:_parton_vrtx}.}
\label{tab:_parton}
\end{table}

\end{widetext}

\subsection{Electromagnetic Response}
\label{subsec:_Ioffe-Larkin}

The electromagnetic response of the system lies at the heart of our discussion. In the vortex theory Eq.~\eqref{eq:_parton_vrtx}, the conserved $\mathrm{U}(1)$ current is given by
\begin{flalign}
J^{\mu}=\frac{\delta\mathcal{L}}{\delta A_{\mu}}=\frac{\mathtt{i}}{2\pi}\varepsilon^{\mu\nu\rho}\partial_{\nu}\tilde{a}_{\rho}.
\label{eq:_dual_current}
\end{flalign}
Consequently, the response function $\Pi^{\mu\nu}$ associated with $J^{\mu}$ (see App.~\ref{App:_Coulomb} for its definition) is determined by the fully dressed gauge-field propagator $\langle\tilde{a}_{\mu}\tilde{a}_{\nu}\rangle$. 

For the problem of a Fermi surface coupled to a gauge field $a_{\mu}=(a_{\tau},\boldsymbol{a})$, it is convenient to work in the Coulomb gauge $\nabla\cdot\boldsymbol{a}=0$~\cite{nayaknfl1,nayaknfl2,HLR1993,lee_nagaosa,nfl_response_1994}. At the level of the random phase approximation (RPA), Eq.~\eqref{eq:_parton_vrtx} leads to the effective theory (see App.~\ref{App:_Coulomb} for our convention) 
\begin{flalign}
\mathcal{S}=\int_{k}\tilde{a}(-k)\frac{\tilde{\Pi}(k)}{2}\tilde{a}(k)-\tilde{a}(-k)\Pi_{\textrm{CS}}(k)A(k)+\ldots
\label{eq:_dual_bosonization}
\end{flalign}
where $k=(\omega,\boldsymbol{k})$ collectively denotes frequency and momentum. The response kernels $\tilde{\Pi}$ and $\Pi_{\textrm{CS}}$ are $2\times2$ matrices expressed in the basis $\tilde{a}=(\tilde{a}_{\tau},\tilde{a}_{T})$ and $A=(A_{\tau},A_{T})$. In our convention, the Chern-Simons kernel takes the form $\Pi_{\textrm{CS}}(k)=-\sigma^{1}|\boldsymbol{k}|/(2\pi)$, where $\sigma^{1}$ denotes the first Pauli matrix. The full gauge kernel $\tilde{\Pi}(k)$ receives contributions from both the fermionic and vortex sectors
\begin{flalign}
\tilde{\Pi}=-\Pi_{\textrm{CS}}\Pi_{f}^{-1}\Pi_{\textrm{CS}}+\Pi_{v},
\label{eq:_Ioffe-Larkin_dual}
\end{flalign}
where $\Pi_{f}$ is the response of the fermionic partons $f$ to $a_{\mu}$ and $\Pi_{v}$ is the response of the vortices $v$ to $\tilde{a}_{\mu}$. The resulting gauge-invariant electromagnetic response of the electron system is then
\begin{flalign}
\Pi^{-1}=-\Pi_{\textrm{CS}}^{-1}\tilde{\Pi}\Pi_{\textrm{CS}}^{-1}=\Pi_{f}^{-1}-\Pi_{\textrm{CS}}^{-1}\Pi_{v}\Pi_{\textrm{CS}}^{-1}.
\label{eq:_Ioffe-Larkin_fv}
\end{flalign}
This general relation holds across the entire phase diagram, provided gauge fluctuations are not too strong.

This expression is equivalent to the Ioffe-Larkin rule Eq.~\eqref{eq:_Ioffe-Larkin_fb} for the original parton construction Eq.~\eqref{eq:_parton_1}, where the duality relation in the chargon sector imposes $\Pi_{v}=-\Pi_{\textrm{CS}}\Pi_{b}^{-1}\Pi_{\textrm{CS}}$. To avoid redundant terminology, we refer to the equivalent expressions under duality transformations, Eq.~\eqref{eq:_Ioffe-Larkin_fb}, Eq.~\eqref{eq:_Ioffe-Larkin_fv}, and Eq.~\eqref{eq:_Ioffe-Larkin_dual}, as the Ioffe-Larkin composition rule~\cite{Ioffe-Larkin}.

From a different perspective, for any fermionic systems with a global U(1) symmetry, one can always introduce a gauge field $\tilde{a}_{\mu}$ to represent the conserved current $J^{\mu}$ via Eq.~\eqref{eq:_dual_current}. Consequently, Eq.~\eqref{eq:_dual_bosonization} can be interpreted as the ``bosonization'' of non-relativistic electrons~\cite{gauge_bosonization_1,gauge_bosonization_2}.

Within the vortex theory framework, the charge disorder operator is represented by the spatial Wilson loop operator of the gauge field $\tilde{a}$
\begin{flalign}
\mathcal{W}_{\mathcal{C}}(\vartheta)=\exp\left(\frac{\mathtt{i}\vartheta}{2\pi}\int_{\mathcal{C}}\tilde{a}\right),
\label{eq:_WL}
\end{flalign}
where $\mathcal{C}$ is a closed loop in real space, and $\vartheta$ is a real-valued parameter. The expectation value $\langle\mathcal{W}_{\mathcal{C}}(\vartheta)\rangle$ can be viewed as a generating functional, with its 2nd cumulant giving the bipartite fluctuations defined in Eq.~\eqref{eq:_BF_def}
\begin{flalign}
\mathcal{F}_{\mathcal{A}}=\lim_{\vartheta\rightarrow0}(-\mathtt{i}\partial_{\vartheta})^{2}\log\langle\mathcal{W}_{\mathcal{C}}(\vartheta)\rangle,
\label{eq:_BF_WL}
\end{flalign}
where $\mathcal{C}=\partial\mathcal{A}$. The large-scale scaling of $\mathcal{F}_{\mathcal{A}}$ is determined by the gauge-field propagator $\langle\tilde{a}_{\mu}\tilde{a}_{\nu}\rangle$ through the Ioffe-Larkin rule Eq.~\eqref{eq:_Ioffe-Larkin_dual}. 


In evaluating Eq.~\eqref{eq:_BF_WL}, one must be very careful and choose a gauge-invariant regularization scheme to handle the UV divergence. For example, directly using the expression in Eq.~\eqref{eq:_Ioffe-Larkin_dual} (under the Coulomb gauge) and setting a real-space cut-off on the integration interval along $\mathcal{C}$ would spoil gauge invariance. We leave the discussion of these technical details to App.~\ref{app:_BF_scaling}.

\subsection{Quantum Criticality}
\label{subsec:_Dyn_deCP}

In this subsection, we clarify the underlying minimal conditions necessary for the universal behaviors of the localization tensor Eq.~\eqref{eq:_QM_log-div} and the bipartite fluctuations Eq.~\eqref{eq:_BF_CFS} to manifest at critical Fermi surfaces.

As will become evident from the examples in the following sections, both Eq.~\eqref{eq:_QM_log-div} and Eq.~\eqref{eq:_BF_CFS} follow directly from the form of the critical response function
\begin{flalign}
\Pi^{-1}=\Pi_{\textrm{FS}}^{-1}+\Pi_{\textrm{CFT}}^{-1},
\label{eq:_Ioffe-Larkin_CFS}
\end{flalign}
where $\Pi_{\textrm{FS}}$ and $\Pi_{\textrm{CFT}}$ denote the response functions of a Fermi surface and a CFT, respectively. This structure appears as a natural consequence of the Ioffe-Larkin rule, Eq.~\eqref{eq:_Ioffe-Larkin_fb} or Eq.~\eqref{eq:_Ioffe-Larkin_fv}. However, it also relies on an additional requirement: that the critical behavior of the chargon (or vortex) sector remains unaffected by its coupling to other low-energy degrees of freedom—a point that requires justification.

At the critical point, the effective theory Eq.~\eqref{eq:_parton_theory} can be schematically written as
\begin{flalign}
\mathcal{L}=\mathcal{L}_{\textrm{FS}}[f,a]+\mathcal{L}_{\textrm{CFT}}[b,A-a]+\mathcal{L}[a]+\mathcal{L}_{\textrm{cp}}[O_{f},O_{b}]+\ldots
\end{flalign}
The two parton sectors are coupled both directly via the term $\mathcal{L}_{\textrm{cp}}$, which involves gauge-invariant operators $O_{f}$ and $O_{b}$, and indirectly through their mutual coupling to the dynamical gauge field $a_{\mu}$. A key task is to determine the fate of these couplings at the quantum critical point.

The leading direct coupling is expected to have the simple form $\mathcal{L}_{\textrm{cp}}\propto O_{f}O_{b}$, with $O_{f}=f^{\dagger}f$ and $O_{b}=|b|^{2}$. It is well known~\cite{sachdevbook} that coupling a bosonic mode to a Fermi surface induces a Landau damping term of the form $\propto\int_{\omega,\boldsymbol{k}}\frac{|\omega|}{|\boldsymbol{k}|}|O_{b}(\omega,\boldsymbol{k})|^{2}$, which is irrelevant if the scaling dimension of $O_{b}$ in the CFT satisfies $\Delta[O_{b}]>3/2$. This condition is firmly established for the 3D XY transition, as confirmed by experiments, Monte Carlo simulations, and conformal bootstrap (see Ref.~\cite{UCt5}). For the continuous transition between a bosonic Laughlin state at $\nu=1/2$ and a superfluid, this condition is also suggested by large-$N$ analysis~\cite{cflfl1,cflfl2}.

To understand the IR behavior of the dynamical gauge field $a_{\mu}$, it is essential to incorporate the effects of Landau damping, which arises from its coupling to the Fermi surface~\cite{senthilmit1,senthilmit2}. The effective action for the transverse component of the gauge field takes the form
\begin{flalign}
\mathcal{S}_{\textrm{eff}}[a]=\int_{\omega,\boldsymbol{k}}\left(\gamma\frac{|\omega|}{|\boldsymbol{k}|}+\chi|\boldsymbol{k}|^{2}+\ldots\right)|a(\omega,\boldsymbol{k})|^{2},
\end{flalign}
where $\chi$ denotes the diamagnetic susceptibility of the $f$-fermions, and the ellipsis includes other contributions such as the polarizability of the $b$-bosons. From this expression, it is clear that gauge fluctuations do not affect the critical properties of the chargon (or vortex) sector. Due to the identical scaling of $\omega$ and $\boldsymbol{k}$ in the CFT, the Landau damping term acts effectively as a Higgs mass term and quenches the dynamics of the gauge field. This suppression of gauge fluctuations lends further support to the validity of the Ioffe-Larkin rule.

In summary, we have reviewed the justification for dynamical decoupling~\cite{senthilmit1,senthilmit2,cflfl2,cflfl3,frac_mit}: both the direct couplings between the two parton sectors and the gauge couplings in $\mathcal{L}_{\textrm{CFT}}[b,A-a]$ are irrelevant at the critical point\footnote{These couplings, however, become important for describing crossovers away from the critical point and for understanding the full phase diagram. For a detailed introduction to the phenomenology, see Ref.~\cite{senthilmit1,senthilmit2}.}. This leads to the important conclusion that correlation functions in the chargon (or vortex) sector retain their CFT form.

We therefore argue that the universal behaviors in Eq.~\eqref{eq:_QM_log-div} and Eq.~\eqref{eq:_BF_CFS} can be understood as direct consequences of two essential ingredients: {\it (1)} the Ioffe-Larkin rule and {\it (2)} dynamical decoupling. These conditions are expected to be satisfied in the examples listed in \tabref{tab:_parton}, which will be discussed in detail in Sec.~\ref{sec:_CFLFL} and Sec.~\ref{sec:_Mott}, and are likely to apply to other similar continuous transitions out of Landau FLs as well.

\section{Continuous CFL-FL Transition}
\label{sec:_CFLFL}

We begin by discussing the critical Fermi surfaces for spinless electrons, focusing on the continuous transition between a Landau FL and a CFL at half-filling. The universal field theory for this transition was first proposed in Ref.~\cite{cflfl2} and subsequently refined in Ref.~\cite{cflfl3}, with particular emphasis on its potential realization in moir{\'e} materials through tuning the electron bandwidth. Experimental evidence supporting the existence of zero-field CFL phases, as well as signatures of the CFL-FL transition, has recently been observed in twisted $\textrm{MoTe}_{2}$ bilayers~\cite{CFLFL_Xu1,CFLFL_Xu2} and in multilayer graphene systems~\cite{CFLFL_Ju}. In parallel, Ref.~\cite{cfl_drude} reported the realization of a CFL-FL transition in a microscopic model of spatially modulated Landau levels by varying the interaction strength, as indicated by the vanishing and nonvanishing behavior of the Drude weight, although whether the transition is truly continuous remains to be determined.


In Sec.~\ref{subsec:_CFLFL_theory}, we briefly discuss the vortex theory that connects Landau FL theory with the HLR theory~\cite{HLR1993}, as well as its relation to the earlier construction~\cite{cflfl2,cflfl3} through fermionic particle-vortex duality. In Sec.~\ref{subsec:_CFLFL_BF}, we clarify the prediction of a universal constant term in Eq.~\eqref{eq:_BF_QH_2} within the CFL phase, as derived from HLR theory, and the universal logarithmic term in Eq.~\eqref{eq:_BF_CFS} at the critical point, and we further examine their connection to the critical resistivity jump described in Eq.~\eqref{eq:_C_jump}.


\subsection{Critical Theories}
\label{subsec:_CFLFL_theory}

Within the vortex theory framework, the critical theory can be described by Eq.~\eqref{eq:_parton_vrtx}, where the fermionic vortices undergo a Chern number changing transition from $\mathsf{C}=0$ to $\mathsf{C}=-2$. This transition can be formulated using two Dirac fermions 
\begin{flalign}
\mathcal{L}_{\textrm{vrtx}}[\psi,\tilde{a}]=\sum_{I=1}^{2}\bar{\psi}_{I}\slashed{\textrm{D}}_{\tilde{a}}\psi_{I}+\frac{1}{2e^{2}}\textrm{d}\tilde{a}\wedge\star\textrm{d}\tilde{a},
\label{eq:_vrtx_f_1}
\end{flalign}
where $\textrm{D}_{\tilde{a}}=\partial-\mathtt{i}\tilde{a}$ denotes the gauge covariant derivative. Note that each Dirac fermion is defined through the Pauli-Villars scheme with another heavy Dirac fermion in the UV. There are two phases depending on the sign of the fermion mass term $m(\bar{\psi}_{1}\psi_{1}+\bar{\psi}_{2}\psi_{2})$. In the case of $\mathsf{C}=0$, as argued in Sec.~\ref{sec:_vrtx}, the Maxwell term in Eq.~\eqref{eq:_vrtx_f_1} becomes important and causes the system to flow back to the Landau FL phase of electrons. In the case of $\mathsf{C}=-2$, the vortex sector in Eq.~\eqref{eq:_parton_vrtx} is given by $\mathcal{L}_{\textrm{vrtx}}=\frac{2\mathtt{i}}{4\pi}\tilde{a}\wedge\textrm{d}\tilde{a}$. After integrating out the gauge field $\tilde{a}$, we have the HLR theory~\cite{HLR1993} for half-filled Landau level 
\begin{flalign}
\mathcal{L}=\mathcal{L}_{\textrm{FS}}[f,a]-\frac{1}{2}\frac{\mathtt{i}}{4\pi}(a-A)\wedge\textrm{d}(a-A),
\label{eq:_HLR}
\end{flalign}
where each $f$-fermion becomes a composite fermion with two flux quanta attached.

For completeness, let us briefly mention the critical theory discussed in Refs.~\cite{cflfl2,cflfl3}, and illustrate its relation to the vortex theory Eq.~\eqref{eq:_parton_vrtx} together with Eq.~\eqref{eq:_vrtx_f_1}. The starting point is the parton construction Eq.~\eqref{eq:_parton_1} described in Sec.~\ref{sec:_vrtx}. The bosonic parton $b$ is assumed to be further fractionalized into two fermions
\begin{flalign}
b(\boldsymbol{x})=f_{1}(\boldsymbol{x})f_{2}(\boldsymbol{x})
\end{flalign}
introducing an additional $\textrm{U}(1)$ gauge field denoted by $\widehat{a}$. One can assign the charge of $A-a$ to $f_{2}$, and put it in a mean-field state with Chern number $\mathsf{C}_{2}=1$. Then a Chern number changing phase transition of $f_{1}$ from $\mathsf{C}_{1}=-1$ to $\mathsf{C}_{1}=1$ will drive the transition of the chargon $b$ from a superfluid state to the bosonic Laughlin state at $\nu=1/2$. The band touching of $f_{1}$ typically involves two massless Dirac fermions $\chi_{1},\chi_{2}$. The full critical theory can be written as 
\begin{flalign}
\mathcal{L}&=\mathcal{L}_{\textrm{FS}}[f,a]+\sum_{I=1}^{2}\bar{\chi}_{I}\slashed{\textrm{D}}_{\widehat{a}}\chi_{I}-\frac{2\mathtt{i}}{4\pi}\widehat{a}\wedge\textrm{d}\widehat{a}\nonumber\\&-\frac{\mathtt{i}}{2\pi}\widehat{a}\wedge\textrm{d}(A-a)-\frac{\mathtt{i}}{4\pi}(A-a)\wedge\textrm{d}(A-a), 
\label{eq:_cflfl_theory_1}
\end{flalign}
where the transition is driven by the fermion mass term $m(\bar{\chi}_{1}\chi_{1}+\bar{\chi}_{2}\chi_{2})$. An important insight from Ref.~\cite{cflfl3} is that a direct second-order phase transition can be protected by translation symmetry and filling constraints. 

As we show in App.~\ref{App:_cflfl_dual}, the critical theories described by Eq.~\eqref{eq:_cflfl_theory_1} and Eq.~\eqref{eq:_parton_vrtx} (together with Eq.~\eqref{eq:_vrtx_f_1}) can be related through fermionic particle-vortex duality~\cite{son2015,dual_review}, where $\psi_{1},\psi_{2}$ can be interpreted as the fermionic vortices of $\chi_{1},\chi_{2}$. Consequently, the transition from the superfluid to the Laughlin state in the chargon sector can be effectively described by the integer QH transition of vortices. The analysis presented in Ref.~\cite{cflfl3} can be carried over for Eq.~\eqref{eq:_vrtx_f_1}, demonstrating that other fermion bilinears are not permitted by translation symmetry under filling constraints.

\subsection{Charge Fluctuations}
\label{subsec:_CFLFL_BF}

Our primary focus is to understand the static structure factor Eq.~\eqref{eq:_SSF} at long wavelengths, which governs the leading-order scaling of bipartite fluctuations $\mathcal{F}$ (see \tabref{tab:_BF_scaling} and App.~\ref{app:_BF_scaling}). Whether a subleading corner contribution appears, as in Eq.~\eqref{eq:_BF_QH_2} or Eq.~\eqref{eq:_BF_CFS}, depends on whether the localization tensor $\mathsf{G}^{ij}$, as defined in Eq.~\eqref{eq:_SSF_expd} and Eq.~\eqref{eq:_SWM_sum}, is finite or logarithmically divergent. In Landau FLs, it is well understood that the leading-order scaling follows $\mathcal{F} \sim L \log(L)$, while the subleading corner contribution is not well-defined due to the slow decay of spatial charge correlations~\cite{uni_corner}. This non-local feature of Landau FLs is directly tied to the power-law divergence of $\mathsf{G}^{ij}$. In the following, we first analyze the CFL phase and then proceed to examine the critical point.

\subsubsection{CFL Phase}
\label{subsubsec:_BF_CFL}

In this subsection, we consider the standard RPA analysis of the HLR theory Eq.~\eqref{eq:_HLR}. For simplicity, we primarily focus on cases with continuous translational and rotational symmetries, as in Refs.~\cite{HLR1993, nfl_response_1994}, with comments on the anisotropic cases provided later. Under the Coulomb gauge, the response function of the $f$-fermions is expressed as $\Pi_{f}(k)=\textrm{diag}(\Pi_{f}^{\tau\tau}(k),\Pi_{f}^{TT}(k))$, where $k=(\omega,\boldsymbol{k})$. At the one-loop level, the longitudinal and transverse components are
\begin{flalign}
\Pi_{f}^{\tau\tau}(k)&=\mathscr{D}_{F}\left(1-\frac{|\omega|}{\sqrt{\omega^{2}+(v_{F}\boldsymbol{k})^{2}}}\right),\nonumber\\\Pi_{f}^{TT}(k)&=-\mathscr{D}_{F}\frac{|\omega|\sqrt{\omega^{2}+(v_{F}\boldsymbol{k})^{2}}-\omega^{2}}{|\boldsymbol{k}|^{2}}.
\label{eq:_FS_response}
\end{flalign}
Here, $v_{F}$ is the fermi velocity, $k_{F}$ denotes the fermi momentum, and $\mathscr{D}_{F}=\frac{k_{F}}{2\pi v_{F}}$ represents the density of states at the fermi level. The vortex sector is in the integer QH state at the filling $\nu=-2$, which has the response function $\Pi_{v}(k)=-2\Pi_{\textrm{CS}}(k)=2\frac{|\boldsymbol{k}|}{2\pi}\sigma^{1}$. The electron density-density correlation is determined using the Ioffe-Larkin rule Eq.~\eqref{eq:_Ioffe-Larkin_fv}. Upon integrating over all frequencies $\omega$, the static structure factor is obtained as follows
\begin{flalign}
&\Pi^{\tau\tau}(\tau\rightarrow0,\boldsymbol{k})=\int\frac{\textrm{d}\omega}{2\pi}\frac{1}{\Pi_{f}^{\tau\tau}(k)^{-1}-\frac{(4\pi)^{2}}{|\boldsymbol{k}|^{2}}\Pi_{f}^{TT}(k)}\nonumber\\&\approx\frac{1-e^{-(|\boldsymbol{k}|\ell_{B})^{2}/2}}{4\pi\ell_{B}^{2}}+\frac{|\boldsymbol{k}|^{3}\log(1/|\boldsymbol{k}|)}{4\pi^{2}k_{F}}+\ldots
\label{eq:_CFL_Pi00}
\end{flalign}
where $\ell_{B}$ denotes the magnetic length scale. 

The first term in Eq.~\eqref{eq:_CFL_Pi00} originates from the inter-Landau-level gapped modes~\cite{GMP1985,GMP1986}. Following a calculation similar to that in Ref.~\cite{uni_corner}, we find that these modes contribute a corner term of $-\textrm{f}(\theta)/(8\pi^{2})$ to the bipartite charge fluctuations, where $\textrm{f}(\theta)$ is the universal angle function defined in Eq.~\eqref{eq:_angle_function}. The second term, proportional to $|\boldsymbol{k}|^{3}\log(1/|\boldsymbol{k}|)$, arises from the gapless modes in the lowest Landau level. This long-wavelength result has been predicted by both the HLR theory~\cite{HLR1993,Read1998} as well as the Son-Dirac theory~\cite{son2015,Dirac_Sci2016}. As shown in App.~\ref{app:_BF_scaling}, the gapless sector contributes only to a boundary-law term. By combining the effects of both gapped and gapless modes, we arrive at the final expression in Eq.~\eqref{eq:_BF_QH} for the configuration shown in \figref{fig:_corner}. Notably, Eq.~\eqref{eq:_BF_QH} holds for both gapped QH insulators~\cite{uni_corner} and gapless CFLs.

It would be highly compelling if Eq.~\eqref{eq:_BF_QH} holds true for all charge insulators, i.e., if the corner contribution is universally determined by the DC Hall conductivity $\sigma_{xy}$. However, unlike at the critical point, deep inside the insulator phase, the gapped modes exhibit short-range correlations and may be sensitive to microscopic details. As we clarify in a follow-up work~\cite{Corner_Geom}, for systems with continuous translational and rotational symmetries, Eq.~\eqref{eq:_BF_QH_2} generally holds, but with a non-universal value of the localization tensor $\mathsf{G}\geq|\sigma_{xy}|/2$. In other words, the coefficient of the universal angle function satisfies a universal lower bound set by $\sigma_{xy}$. In Ref.~\cite{Corner_Geom}, we explain this bound and its saturation from the perspective of many-body quantum geometry\footnote{For gapped insulators, $\mathsf{G}^{ij}=\mathsf{G}\delta^{ij}$ serves as the many-body quantum metric~\cite{Corner_Geom}, defined by adiabatic flux insertion (or twisted boundary conditions), where $i=x,y$ denotes the spatial indices.} (see also App.~\ref{App:_Optical_Bounds}). It turns out that $\mathsf{G}=|\sigma_{xy}|/2$ is guaranteed under specific conditions, such as Galilean invariance and wavefunction holomorphicity, which are indeed satisfied by many examples. In fact, our predicted formula Eq.~\eqref{eq:_BF_QH} has already been numerically verified by Monte Carlo simulations in Ref.~\cite{CFLEE3} using the Rezayi-Read wavefunction, considering CFLs for fermions at $\nu=1/2$ and $\nu=1/4$, as well as CFLs for bosons at $\nu=1$ and $\nu=1/3$ fillings.

More generally, the system may lack continuous translational and rotational symmetries, as in the models with spatially modulated periodic potentials discussed in Refs.~\cite{tmdcfl1,tmdcfl2,cfl_drude}, which concern the half-filled Chern band in mori\'{e} materials. In such cases, the corner contribution may depend on the corner angle, subregion orientation, and other microscopic details. Nevertheless, in Ref.~\cite{Corner_Geom}, we demonstrate that the corner contribution can be expressed as an infinite series of orientation-resolved universal angle functions, generalizing the formula Eq.~\eqref{eq:_angle_function}, along with their non-universal coefficients. However, these details are beyond the scope of this work.


\subsubsection{Critical Point}
\label{subsubsec:_BF_CFLFL}

Next, we turn our attention to the critical point, where the analysis becomes slightly more involved. The QH transition in the vortex sector (i.e., the chargon sector) is presumably described by a CFT. As argued in Refs.~\cite{cflfl2,cflfl3}, the CFT sector and the $f$-fermion sector are dynamically decoupled precisely at the critical point—a reasoning we have reviewed in Sec.~\ref{subsec:_Dyn_deCP}.


As a result, the critical response in the vortex (or chargon) sector retains its CFT form\footnote{In terms of space-time components, the current-current correlation has the form $\Pi_{\textrm{CFT}}^{\mu\nu}(k)=\sigma_{xx}|k|(\delta_{\mu\nu}-\frac{k_{\mu}k_{\nu}}{|k|^{2}})+\sigma_{xy}\varepsilon_{\mu\nu\rho}k_{\rho}$.}. Under the Coulomb gauge, it is expressed as
\begin{flalign}
\Pi_{\textrm{CFT}}(k)=\left(\begin{array}{cc}
\frac{|\boldsymbol{k}|^{2}}{\omega}\mathsf{F}_{L}(\frac{\omega}{|\boldsymbol{k}|}) & \sigma_{xy}|\boldsymbol{k}|\\
\sigma_{xy}|\boldsymbol{k}| & -\omega\mathsf{F}_{T}(\frac{\omega}{|\boldsymbol{k}|})
\end{array}\right),
\label{eq:_CFT_response_1}
\end{flalign}
where the dimensionless scaling functions of the longitudinal and transverse components are given by
\begin{flalign}
\mathsf{F}_{L}(\lambda)&=\sigma_{xx}\frac{\lambda}{\sqrt{1+\lambda^{2}}},\nonumber\\\mathsf{F}_{T}(\lambda)&=\sigma_{xx}\frac{\sqrt{1+\lambda^{2}}}{\lambda}.
\label{eq:_CFT_response_2}
\end{flalign}
Here, $\sigma_{xx}$ and $\sigma_{xy}$ represent the universal longitudinal and Hall conductivities of the CFT. Their values in $\Pi_{v}(k)$ for vortices and in $\Pi_{b}(k)$ for chargons are related by
\begin{flalign}
\sigma_{xx}^{b}&=\frac{1}{(2\pi)^{2}}\frac{\sigma_{xx}^{v}}{(\sigma_{xx}^{v})^{2}+(\sigma_{xy}^{v})^{2}},\nonumber\\\sigma_{xy}^{b}&=\frac{1}{(2\pi)^{2}}\frac{-\sigma_{xy}^{v}}{(\sigma_{xx}^{v})^{2}+(\sigma_{xy}^{v})^{2}}.
\label{eq:_conductivity_dual}
\end{flalign}

According to the Ioffe-Larkin rule Eq.~\eqref{eq:_Ioffe-Larkin_fb} or Eq.~\eqref{eq:_Ioffe-Larkin_fv}, the static structure factor can be calculated as follows
\begin{widetext}
\begin{flalign}
\Pi^{\tau\tau}(\tau\rightarrow0,\boldsymbol{k})=\int\frac{\textrm{d}\omega}{2\pi}\frac{|\boldsymbol{k}|^{2}\Pi_{f}^{\tau\tau}(k)(\mathsf{F}_{L}^{b}(\frac{\omega}{|\boldsymbol{k}|})(\Pi_{f}^{TT}(k)-\omega\mathsf{F}_{T}^{b}(\frac{\omega}{|\boldsymbol{k}|}))+\omega(\sigma_{xy}^{b})^{2})}{(\Pi_{f}^{TT}(k)-\omega\mathsf{F}_{T}^{b}(\frac{\omega}{|\boldsymbol{k}|}))(|\boldsymbol{k}|^{2}\mathsf{F}_{L}^{b}(\frac{\omega}{|\boldsymbol{k}|})+\omega\Pi_{f}^{\tau\tau}(k))+\omega|\boldsymbol{k}|^{2}(\sigma_{xy}^{b})^{2}}=\frac{\sigma_{xx}^{b}}{\pi}|\boldsymbol{k}|^{2}\log(1/|\boldsymbol{k}|)+\ldots
\label{eq:_CFLFL_SSF}
\end{flalign}
\end{widetext}
Here, $\Pi_{f}^{\tau\tau}(k)$ and $\Pi_{f}^{TT}(k)$ are the components of the response function of $f$-fermions given in Eq.~\eqref{eq:_FS_response}. The scaling functions $\mathsf{F}_{L}^{b}(k)$ and $\mathsf{F}_{T}^{b}(k)$ are provided by  Eq.~\eqref{eq:_CFT_response_2}, with the coefficient $\sigma_{xx}^{b}$. Evaluating the frequency integral in Eq.~\eqref{eq:_CFLFL_SSF} is not an easy task. In App.~\ref{App:_CFS_SSF}, we present two methods for extracting the long-wavelength behavior $|\boldsymbol{k}|^{2}\log(1/|\boldsymbol{k}|)$. By retaining only the leading-order term of the integrand in the small-$\boldsymbol{k}$ expansion, the $\omega$-integral can be performed analytically. Additionally, we have numerically evaluated the full expression in Eq.~\eqref{eq:_CFLFL_SSF}. Remarkably, at small $|\boldsymbol{k}|/k_{F}$, the numerical result aligns closely with the analytical expression.

The Fourier transform of Eq.~\eqref{eq:_CFLFL_SSF} yields the power-law spatial correlation given by Eq.~\eqref{eq:_power-law_Pi00}, with an exponent $\alpha = 4$. The overall coefficient $C_{0}$ is given by the universal constant $C_{\rho}=2(\sigma_{xx}^{b})/\pi^{2}$, which identifies the current central charge of the CFT describing the chargon sector. By examining the integrand in Eq.~\eqref{eq:_CFLFL_SSF} and sequentially taking the limits $\omega\rightarrow0$ and then $\boldsymbol{k}\rightarrow0$, we can demonstrate the vanishing compressibility. Moreover, the vanishing Drude weight can be confirmed using the Ioffe-Larkin rule Eq.~\eqref{eq:_Ioffe-Larkin_fb} or Eq.~\eqref{eq:_Ioffe-Larkin_fv}. As a result, we identify an incompressible state with vanishing Drude weight, characterized by a logarithmically divergent localization tensor as described by Eq.~\eqref{eq:_QM_log-div}, despite the absence of conformal symmetry.

According to Refs.~\cite{Dirac_log, wulog, chengdisop, uni_corner} (see also App.~\ref{app:_BF_scaling}), for the subregion depicted in \figref{fig:_corner}, the bipartite charge fluctuations follow the CFT-like expression in Eq.~\eqref{eq:_BF_CFS}. Similar to the first term in Eq.~\eqref{eq:_CFL_Pi00}, Eq.~\eqref{eq:_CFLFL_SSF} also includes a $|\boldsymbol{k}|^2$ contribution from gapped modes, implying that the localization tensor contains finite subleading terms in addition to the divergence in Eq.~\eqref{eq:_QM_log-div}. These modes contribute to a corner term of the form $(\textrm{const})\textrm{f}(\theta)$ that does not scale with $L$, similar to the subleading contribution in Eq.~\eqref{eq:_BF_QH_2}. The coefficient of this term depends on $\sigma_{xx}^{b}$ and $\sigma_{xy}^{b}$. However, isolating this subleading constant from the $\log(L)$ scaling in experiments or numerical simulations can be challenging. Theoretically, this constant term is also sensitive to the UV cutoff.

One may wonder whether the universal number $C_{\rho}$ can also be probed through other experimental observables, such as transport measurements. A promising candidate is the predicted resistivity jump at the critical point~\cite{cflfl3}, which is entirely determined by universal data from the CFT sector. The key argument~\cite{cflfl3} leading to this universal resistivity jump is based on the observation that $\Pi^{\tau j}(\mathtt{i}\omega,\boldsymbol{k}\rightarrow0)=0$, where $j=x,y$, holds for both the $f$-fermion and the $b$-chargon sectors. Considering the spatial components of the Ioffe-Larkin rule Eq.~\eqref{eq:_Ioffe-Larkin_fb} or Eq.~\eqref{eq:_Ioffe-Larkin_fv}, the total resistivity is given by $\rho_{ij}=\rho^{f}_{ij}+\rho^{b}_{ij}$. As one approaches the critical point from the Landau FL phase, additional contributions to the longitudinal and Hall components arise from the gapless degrees of freedom in the CFT:
\begin{flalign}
\Delta\rho_{xx}&=\rho_{xx}^{b}=\frac{\sigma_{xx}^{b}}{(\sigma_{xx}^{b})^{2}+(\sigma_{xy}^{b})^{2}},\nonumber\\\Delta\rho_{xy}&=\rho_{xy}^{b}=\frac{-\sigma_{xy}^{b}}{(\sigma_{xx}^{b})^{2}+(\sigma_{xy}^{b})^{2}}.
\end{flalign}
These expressions allow one to solve for $\sigma_{xx}^{b}$ and express the universal coefficient $C_{\rho}=2(\sigma_{xx}^{b})/\pi^{2}$ in Eq.~\eqref{eq:_BF_CFS} as a function of $\Delta\rho_{xx}$ and $\Delta\rho_{xy}$, as shown in Eq.~\eqref{eq:_C_jump}. Note that the resistivity tensor $\Delta\rho_{ij}$ is a scaling function of $\omega/T$ (i.e., frequency over temperature), and Eq.~\eqref{eq:_C_jump} should be interpreted as a relation in the limit $T\rightarrow0$. The values of $\Delta\rho_{xx}$ and $\Delta\rho_{xy}$, computed via large-$N$ expansions, are available in Refs.~\cite{wufisher, cflfl1}, and could potentially be determined more precisely using conformal bootstrap techniques~\cite{bootstrapRMP,bootstrapRMP2}. 

We conclude this section by emphasizing that the predicted universal behaviors in Eq.~\eqref{eq:_QM_log-div} and Eq.~\eqref{eq:_BF_CFS} are robust and remain independent of the specific form of the Fermi-surface response function $\Pi_{f}$. For illustration purposes, we have used the special case given by Eq.~\eqref{eq:_FS_response} in the integral Eq.~\eqref{eq:_CFLFL_SSF}. However, in general, $\Pi_{f}$ is determined by the electron band structure and may exhibit anisotropy and sensitivity to UV details. Despite this, Eq.~\eqref{eq:_QM_log-div} and Eq.~\eqref{eq:_BF_CFS} remain valid. This is because, for the purpose of evaluating corner charge fluctuations, one essentially has $\Pi^{-1}=\Pi_{f}^{-1}+\Pi_{b}^{-1}\approx\Pi_{b}^{-1}$, where $\Pi_{b}$ takes the CFT expression Eq.~\eqref{eq:_CFT_response_1} due to the emergent conformal symmetry in the vortex/chargon sector at criticality. This result follows as a general consequence of the Ioffe-Larkin rule and the distinct scaling behaviors of $\Pi_{b}$ and $\Pi_{f}$ at the critical point.

\section{Continuous Mott Transition}
\label{sec:_Mott}

In this section, we now turn to spin-$1/2$ electrons, which enable the definition of both charge and spin fluctuations within a subsystem. We examine two theoretical constructions of continuous Mott transitions at half-filling that preserve time-reversal symmetry. Due to their technical similarities to the CFL-FL transition, some of the results from Sec.~\ref{sec:_CFLFL} are applicable here. We begin with the original proposal~\cite{lee2005,senthilmit1,senthilmit2}, which was motivated by the Mott organic compound $\kappa$-$\textrm{(ET)}_{2}\textrm{Cu}_{2}\textrm{(CN)}_{3}$~\cite{mit_organic}, and also discuss a generalized multicritical vortex theory in this context. We then turn to a modified theory~\cite{frac_mit} proposed for another candidate material, the TMD mori\'{e} bilayer $\textrm{MoTe}_{2}/\textrm{WSe}_{2}$, which exhibits an anomalously large critical resistivity in the experiment~\cite{tmdmit1}. 


\subsection{Charge Fluctuations}

To cause the Fermi surface of electrons to disappear abruptly in a continuous manner, spin-charge separation is necessary, leaving a neutral Fermi surface on the insulator side. The original theoretical proposal~\cite{lee2005,senthilmit1,senthilmit2} was based on the parton construction
\begin{flalign}
c_{\sigma}(\boldsymbol{x})=b(\boldsymbol{x})f_{\sigma}(\boldsymbol{x}), 
\label{eq:_parton2}
\end{flalign}
where each electron $c_{\sigma}$ is fractionalized into a spinless bosonic chargon $b$, which carries the electric charge, and a charge-neutral fermionic spinon $f_{\sigma}$, which carries the spin quantum number $\sigma=\uparrow,\downarrow$. There is a dynamical $\textrm{U}(1)$ gauge field $a_{\mu}$ that couples $b$ and $f$. After introducing the dual bosonic vortices $\varphi$ of the chargons $b$, the critical theory is described by Eq.~\eqref{eq:_parton_vrtx}, where $\mathcal{L}_{\textrm{FS}}[f_{\sigma},a]$ describes a Fermi surface of spinons. The continuous Mott transition is driven by a superfluid-to-Mott-insulator transition of the chargons. In the dual description, this is captured by the condensation of vortices, governed by 
\begin{flalign}
\mathcal{L}_{\textrm{vrtx}}[\varphi,\tilde{a}]=|\textrm{D}_{\tilde{a}}\varphi|^{2}+r|\varphi|^{2}+u|\varphi|^{4},
\end{flalign}
where $\textrm{D}_{\tilde{a}}=\partial-\mathtt{i}\tilde{a}$ is the gauge covariant derivative. A single tuning parameter $r\sim(g-g_{c})$, with $g$ denoting the electronic bandwidth, controls the transition. For $r>0$, the vortex field $\varphi$ is trivially gapped, corresponding to the Landau FL phase of electrons. For $r<0$, vortex condensation Higgses the gauge field $\tilde{a}$, driving the system into a spin-liquid Mott insulator characterized by a gapless spinon Fermi surface. The expectation for the distinct scalings of bipartite fluctuations in the two phases was briefly mentioned in Ref.~\cite{FSEE_swingle2}.

An important feature of the critical theory is the dynamical decoupling between the chargon (or vortex) sector and the spinon sector, as the conditions outlined in Sec.~\ref{subsec:_Dyn_deCP} are satisfied by the 3D XY fixed point~\cite{senthilmit1,senthilmit2}. As a result, the zero-temperature response function $\Pi_{b}$ of the chargons follows the CFT form in Eq.~\eqref{eq:_CFT_response_1}, with $\sigma_{xx}^{b}$ being the critical conductivity of the 3D XY universality class, and $\sigma_{xy}^{b}=0$. In the dual picture, the response function of vortices $\Pi_{v}$ is also of the same form Eq.~\eqref{eq:_CFT_response_1}, with the coefficients $\sigma_{xx}^{v}=1/(4\pi^{2}\sigma_{xx}^{b})$ and $\sigma_{xy}^{v}=0$. Assuming the $f$-fermions remain in their mean-field state with response function $\Pi_{f}$ given by Eq.~\eqref{eq:_FS_response}, the Ioffe-Larkin rule Eq.~\eqref{eq:_Ioffe-Larkin_fv} or Eq.~\eqref{eq:_Ioffe-Larkin_fb} yields the leading long-wavelength behavior of the static structure factor
\begin{flalign}
\Pi^{\tau\tau}(\tau\rightarrow0,\boldsymbol{k})&=\int\frac{\textrm{d}\omega}{2\pi}\frac{1}{\Pi_{f}^{\tau\tau}(k)^{-1}+\frac{\omega}{|\boldsymbol{k}|^{2}}\mathsf{F}_{L}^{b}(\frac{\omega}{|\boldsymbol{k}|})^{-1}}\nonumber\\&=\frac{\sigma_{xx}^{b}}{\pi}|\boldsymbol{k}|^{2}\log(1/|\boldsymbol{k}|)+\ldots
\label{eq:_Mott_SSF}
\end{flalign}
Since the Hall response vanishes, the expression for the Mott quantum criticality in Eq.~\eqref{eq:_Mott_SSF} takes a simpler form compared to that for the CFL-FL transition in Eq.~\eqref{eq:_CFLFL_SSF}. A common feature of both transitions is that they describe incompressible states with vanishing Drude weight, characterized by a static structure factor exhibiting CFT-like behavior with an overall coefficient set by the universal conductivity $\sigma_{xx}^{b}$ of the chargons.

For the subsystem geometry shown in \figref{fig:_corner}, the bipartite charge fluctuations follow Eq.~\eqref{eq:_BF_CFS}, and feature a universal corner contribution proportional to the current central charge of the 3D XY fixed point
\begin{flalign}
C_{\rho}=\frac{2}{\pi^{2}}\sigma_{xx}^{b}=C_{J}^{(\textrm{XY})}
\label{eq:_CJ_mit1}
\end{flalign}
 Due to time-reversal invariance, the universal coefficient is directly tied to the longitudinal resistivity jump at the critical point. With a vanishing Hall response, the total longitudinal resistivity is given by the Ioffe-Larkin rule $\rho_{xx}=\rho_{xx}^{f}+\rho_{xx}^{b}$, with $\rho_{xx}^{f}$ and $\rho_{xx}^{b}$ denoting the spinon and chargon contributions, respectively. At zero temperature, the insulating phase features $\rho_{xx}^{b} = +\infty$ due to the chargon gap, leading to diverging total resistivity. In contrast, in the metallic phase, the chargon superfluid leads to $\rho_{xx}^{b} = 0$, so $\rho_{xx}$ is entirely determined by the spinon sector. At the critical point, the universal chargon contribution induces a finite jump $\Delta\rho_{xx}(\omega/T)$ in the longitudinal resistivity~\cite{senthilmit2}. In the limit $\omega/T \to \infty$, this jump is given by $\Delta\rho_{xx} = 2/(\pi^{2} C_{J}^{(\textrm{XY})})$, establishing a direct link between the critical resistivity and the universal corner term in bipartite charge fluctuations. In clean systems with negligible spinon resistivity $\rho_{xx}^{f} \approx 0$, the total critical resistivity is well approximated by $\Delta\rho_{xx}$.


\subsection{Spin Fluctuations}
\label{subsec:_Mott_sc}

In addition to the $\textrm{U}(1)$ charge symmetry, the system also possesses a $\textrm{U}(1)$ spin symmetry associated with the conservation of the third component of spin. A deeper understanding can be achieved by incorporating both charge and spin vortices $\varphi^{c},\varphi^{s}$ in strongly correlated metals. The full theory is introduced as follows
\begin{widetext}
\begin{flalign}
\mathcal{L}=\;&\mathcal{L}_{\textrm{FS}}[f,a^{c},a^{s}]+\frac{\mathtt{i}}{2\pi}\tilde{a}^{c}\wedge\textrm{d}(A^{c}-a^{c})+\frac{\mathtt{i}}{2\pi}\tilde{a}^{s}\wedge\textrm{d}(A^{s}-a^{s})+\frac{1}{2e_{c}^{2}}\textrm{d}\tilde{a}^{c}\wedge\star\textrm{d}\tilde{a}^{c}+\frac{1}{2e_{s}^{2}}\textrm{d}\tilde{a}^{s}\wedge\star\textrm{d}\tilde{a}^{s}\nonumber\\&+|(\partial-\mathtt{i}\tilde{a}^{c})\varphi^{c}|^{2}+r_{c}|\varphi^{c}|^{2}+u_{c}|\varphi^{c}|^{4}+|(\partial-\mathtt{i}\tilde{a}^{s})\varphi^{s}|^{2}+r_{s}|\varphi^{s}|^{2}+u_{s}|\varphi^{s}|^{4}+\ldots.
\label{eq:_parton_vrtx_sc}
\end{flalign}
\end{widetext}
Here, the background fields $A^{c}$ and $A^{s}$ are introduced to keep track of the global charge and spin $\textrm{U}(1)$ symmetries, respectively. The spinon field $f=(f_{\uparrow},f_{\downarrow})$ forms a Fermi surface and is minimally coupled to emergent $\textrm{U}(1)$ gauge fields $a^{c}$ and $a^{s}$ through the covariant derivative $\textrm{D}=\partial\sigma^{0}-\mathtt{i}a^{c}\sigma^{0}-\mathtt{i}a^{s}\sigma^{3}$, where $\sigma^{\mu}$ denotes the vector of Pauli matrices. The fluxes of the $\textrm{U}(1)$ gauge fields $\tilde{a}^{c}$ and $\tilde{a}^{s}$ represent the densities of bosonic partons that carry the global charge and spin $\textrm{U}(1)$ quantum numbers, respectively. The parton construction underlying Eq.~\eqref{eq:_parton_vrtx_sc} is reminiscent of the $\textrm{U}(2)$ gauge theory framework discussed in Ref.~\cite{zouMott}. There is no a priori relation between the two sets of couplings $e, r, u$ associated with the charge and spin vortex sectors.

Based on the vortex theory Eq.~\eqref{eq:_parton_vrtx_sc}, two gauge-invariant Wilson loop operators can be defined:
\begin{flalign}
\mathcal{W}_{\mathcal{C}}^{c}(\vartheta)&=\exp\left(\frac{\mathtt{i}\vartheta}{2\pi}\int_{\mathcal{C}}\tilde{a}^{c}\right),\nonumber\\\mathcal{W}_{\mathcal{C}}^{s}(\vartheta)&=\exp\left(\frac{\mathtt{i}\vartheta}{2\pi}\int_{\mathcal{C}}\tilde{a}^{s}\right),
\end{flalign}
which represent the charge-$\textrm{U}(1)$ and spin-$\textrm{U}(1)$ disorder operators associated with the subsystem $\mathcal{A}$, where $\mathcal{C}=\partial\mathcal{A}$. The charge and spin fluctuations $\mathcal{F}_{\mathcal{A}}^{c}$ and $\mathcal{F}_{\mathcal{A}}^{s}$ are then obtained using $\mathcal{W}_{\mathcal{C}}^{c}$ and $\mathcal{W}_{\mathcal{C}}^{s}$, as described in Eq.~\eqref{eq:_BF_WL}.

The vortex theory described by Eq.~\eqref{eq:_parton_vrtx_sc} is a multi-critical theory governed by two coupling constants, $r_{c}$ and $r_{s}$. The Mott transition discussed earlier corresponds to the situation where the spin vortex $\varphi^{s}$ remains gapped, while the condensation of the charge vortex $\varphi^{c}$ is dual to the 3D XY transition. When both $\varphi^{c}$ and $\varphi^{s}$ are gapped, the system stays in the ordinary metallic phase. In contrast, if $\varphi^{c}$ condenses, the system enters a Mott insulating phase characterized by gapless spinons coupled to a deconfined gauge field $a^{c}$, with $\tilde{a}^{c}$ gapped out by the Higgs mechanism. Conversely, if $\varphi^{c}$ remains gapped while $\varphi^{s}$ condenses, the transition leads to an algebraic charge liquid, featuring power-law charge correlations. The expected behaviors of charge and spin bipartite fluctuations are summarized in \tabref{tab:_BF_Mott_sc}. Regarding the final fate in the IR, there is an important distinction between the two phases where a single vortex condenses. In the $\varphi^{s}$-condensed case, the gauge field $a^{s}$ eventually drives a pairing instability of the Fermi-surface state, resulting in a superconducting ground state. Nevertheless, the intermediate charge liquid phase should still be observable within a finite energy window.

\begin{table}
\begin{tabular}{|c|c|c|}
\hline 
 & $\varphi^{s}$-gapped & $\varphi^{s}$-condensed\tabularnewline
\hline 
$\varphi^{c}$-gapped & $(L\log L,L\log L)$ & $(L\log L,L)$\tabularnewline
\hline 
$\varphi^{c}$-condensed & $(L,L\log L)$ & $(L,L)$\tabularnewline
\hline 
\end{tabular}
\caption{Leading-order scalings of charge and spin bipartite fluctuations $(\mathcal{F}^{c},\mathcal{F}^{s})$ across the transitions governed by the multicritical vortex theory Eq.~\eqref{eq:_parton_vrtx_sc}
involving charge and spin vortices $\varphi^{c}$ and $\varphi^{s}$.
At the critical point associated with each vortex condensation, the bipartite fluctuations exhibit a universal corner contribution proportional to the current central charge of the 3D XY fixed point.
}
\label{tab:_BF_Mott_sc}
\end{table}

When both $\varphi^{c}$ and $\varphi^{s}$ are condensed, the $f$-fermions become coupled to two $\textrm{U}(1)$ gauge fields
\begin{equation}
\mathcal{L}=\mathcal{L}_{\textrm{FS}}[f,a^{c},a^{s}]+\frac{1}{2e_{c}^{2}}\textrm{d}a^{c}\wedge\star\textrm{d}a^{c}+\frac{1}{2e_{s}^{2}}\textrm{d}a^{s}\wedge\star\textrm{d}a^{s}.
\end{equation}
There is a competition between the effects of $a^{s}$, which drives a pairing instability, and $a^{c}$, which suppresses the tendency towards pairing. A perturbative RG analysis in Ref.~\cite{zouMott} shows that the IR fate of the system depends on the UV coupling constants. If $e_{c}/e_{s}<1$, the Fermi-surface state becomes unstable and favors pairing. Conversely, if $e_{c}/e_{s}>1$, the system flows to a stable non-Fermi liquid fixed point. The case where $e_{c}/e_{s}=1$ resembles the behavior of Landau FLs, with the stability of the Fermi-surface state depending on whether the pairing interaction is repulsive or attractive in the UV.

\subsection{Charge Fractionalization}

Another type of vortex theory was introduced in Ref.~\cite{frac_mit}, inspired by the experimental observation of a continuous Mott transition in the TMD heterobilayer $\textrm{MoTe}_{2}/\textrm{WSe}_{2}$~\cite{tmdmit1}. In this theory, two bosonic chargons $b_{\uparrow}$ and $b_{\downarrow}$ are introduced for the two spin/valley quantum numbers, defined as
\begin{flalign}
c_{\uparrow}(\boldsymbol{x})=b_{\uparrow}(\boldsymbol{x})f_{\uparrow}(\boldsymbol{x}),\quad c_{\downarrow}(\boldsymbol{x})=b_{\downarrow}(\boldsymbol{x})f_{\downarrow}(\boldsymbol{x}), \label{eq:_parton3}
\end{flalign}
where $f_{\sigma}$ represents the fermionic spinon. Because of time-reversal symmetry, the condensation of both $b_{\uparrow}$ and $b_{\downarrow}$ occurs simultaneously, leading to the metal-insulator transition. The novelty of this theory lies in the filling factor of the chargons. When the electron $c$ is at half-filling, both $b_{\uparrow}$ and $b_{\downarrow}$ are also at half-filling, in contrast to the integer filling of $b$ in the construction given by Eq.~\eqref{eq:_parton2}. In this case, the Lieb-Shultz-Matthis (LSM) theorem~\cite{LSM,hastings} dictates that the Mott insulator phase of each chargon cannot be a trivial insulator. Instead, it must either be a topological order or form a density wave that spontaneously breaks the translation symmetry. In both cases, the critical point exhibits charge fractionalization and leads to anomalously large critical resistivity~\cite{frac_mit}.  For simplicity, in this section, we focus on the case of topological order for illustration purposes.

We adopt the critical theory from Ref.~\cite{frac_mit} as follows
\begin{widetext}
\begin{flalign}
\mathcal{L}=\;&\mathcal{L}_{\textrm{FS}}[f_{\uparrow},a^{\uparrow}]+\mathcal{L}_{\textrm{FS}}[f_{\downarrow},a^{\downarrow}]+\frac{\mathtt{i}}{2\pi}\tilde{a}^{\uparrow}\wedge\textrm{d}(A^{c}+A^{s}-a^{\uparrow})+\frac{\mathtt{i}}{2\pi}\tilde{a}^{\downarrow}\wedge\textrm{d}(A^{c}-A^{s}-a^{\downarrow})+\frac{1}{2e^{2}}\textrm{d}\tilde{a}^{\uparrow}\wedge\star\textrm{d}\tilde{a}^{\uparrow}\nonumber\\&+\frac{1}{2e^{2}}\textrm{d}\tilde{a}^{\downarrow}\wedge\star\textrm{d}\tilde{a}^{\downarrow}+|(\partial-\mathtt{i}N\tilde{a}^{\uparrow})\varphi^{\uparrow}|^{2}+|(\partial-\mathtt{i}N\tilde{a}^{\downarrow})\varphi^{\downarrow}|^{2}+r(|\varphi^{\uparrow}|^{2}+|\varphi^{\downarrow}|^{2})+u|\varphi^{\uparrow}|^{4}+u|\varphi^{\downarrow}|^{4}+\ldots.
\label{eq:_parton_vrtx_ud}
\end{flalign}
\end{widetext}
The background fields $A^{c}$ and $A^{s}$ are defined in the same way as in Eq.~\eqref{eq:_parton_vrtx_sc}. The two emergent $\textrm{U}(1)$ gauge fields $a^{\uparrow}$ and $a^{\downarrow}$ arise from the gauge redundancies in the parton construction Eq.~\eqref{eq:_parton3}. The fluxes of the $\textrm{U}(1)$ gauge fields $\tilde{a}^{\uparrow}$ and $\tilde{a}^{\downarrow}$ represent the densities of the chargons $b_{\uparrow}$ and $b_{\downarrow}$ respectively, while the vortex excitations $\varphi^{\uparrow},\varphi^{\downarrow}$ are charged under $\tilde{a}^{\uparrow}$ and $\tilde{a}^{\downarrow}$. In Eq.~\eqref{eq:_parton_vrtx_ud}, the theory essentially contains two decoupled sectors labeled by the quantum numbers $\uparrow$ and $\downarrow$, as inter-valley couplings involve high-energy processes requiring large momentum transfer. Due to time-reversal symmetry, the two sectors are identical, meaning that the fermionic partons $f_{\uparrow},f_{\downarrow}$ are in the same mean-field state, with identical coupling constants in both sectors. Because of the $\pi$ flux background of $\tilde{a}^{\uparrow}$ (and $\tilde{a}^{\downarrow}$) per unit cell, the vortex dynamics are frustrated. Without breaking lattice translation symmetry, only $N$-vortex bound states can condense, where $N$ must be an even integer. In Eq.~\eqref{eq:_parton_vrtx_ud}, the bosonic field $\varphi^{\uparrow}$ and $\varphi^{\downarrow}$ represent these $N$-vortex bound states. 

The condensation of both $\varphi^{\uparrow}$ and $\varphi^{\downarrow}$ is controlled by a single coupling constant $r$, which is related to the bandwidth of the electrons. In the Mott insulator, each spin/valley sector forms a $\mathbb{Z}_{N}$ topological order. Following an analysis similar to that in Sec.~\ref{subsec:_Dyn_deCP}, it can be shown that Landau damping effects are irrelevant at the 3D XY* transition. As before, the vortex sector associated with each spin (or valley) is dynamically decoupled from the spinon Fermi surface. At both the Mott insulating phase and the critical point, the charge carriers are anyons associated with the $\mathbb{Z}_{N}$ topological order, each carrying a fractional electric charge of $e_{*}=e/N$. 

The charge and spin disorder operators are represented by the Wilson loop operators
\begin{flalign}
\mathcal{W}_{\mathcal{C}}^{c}(\vartheta)&=\exp\left(\frac{\mathtt{i}\vartheta}{2\pi}\int_{\mathcal{C}}(\tilde{a}^{\uparrow}+\tilde{a}^{\downarrow})\right),\nonumber\\\mathcal{W}_{\mathcal{C}}^{s}(\vartheta)&=\exp\left(\frac{\mathtt{i}\vartheta}{2\pi}\int_{\mathcal{C}}(\tilde{a}^{\uparrow}-\tilde{a}^{\downarrow})\right),
\end{flalign}
which define the charge and spin bipartite fluctuations $\mathcal{F}_{\mathcal{A}}^{c}$ and $\mathcal{F}_{\mathcal{A}}^{s}$ through Eq.~\eqref{eq:_BF_WL}, where $\mathcal{C}=\partial\mathcal{A}$. The scaling behaviors of $\mathcal{F}^{c}$ can be easily seen to follow $L\log L$ and $ L$ by using the gauge-field propagators $\langle\tilde{a}^{\uparrow}\tilde{a}^{\uparrow}\rangle$ and $\langle\tilde{a}^{\downarrow}\tilde{a}^{\downarrow}\rangle$ before and after the vortex condensation. At the critical point, an interesting universal corner contribution appears, with a coefficient that deviates from Eq.~\eqref{eq:_CJ_mit1}. The self-energy of $\tilde{a}^{\uparrow}$ is dominated by the loop corrections from $\varphi^{\uparrow}$, which is proportional to $N^{2}$. Consequently, the two-point function of the conserved current $J^{\uparrow}=\frac{\mathtt{i}}{2\pi}\star\textrm{d}\tilde{a}^{\uparrow}$ at the XY* fixed point takes the form
\begin{flalign}
\left\langle J_{\mu}^{\uparrow}(x)J_{\nu}^{\uparrow}(0)\right\rangle =\frac{C_{J}^{(\textrm{XY})}}{N^{2}}\frac{1}{|x|^{4}}\left(\delta_{\mu\nu}-\frac{2x_{\mu}x_{\nu}}{|x|^{2}}\right),    
\end{flalign}
where we still use $C_{J}^{(\textrm{XY})}$ to denote the current central charge at the XY fixed point. Identical expressions hold for $\tilde{a}^{\downarrow}$ and $J^{\downarrow}$. As a result, the charge fluctuations $\mathcal{F}^{c}$ again follow Eq.~\eqref{eq:_BF_CFS}, now with the universal coefficient
\begin{flalign}
C_{\rho}=\frac{2C_{J}^{(\textrm{XY})}}{N^{2}}.
\end{flalign}
This value of $C_{\rho}$ allows one to determine the universal resistivity jump $\Delta\rho_{xx}=N^{2}/(\pi^{2}C_{J}^{(\textrm{XY})})$, as given in Eq.~\eqref{eq:_C_jump}. It is argued in Ref.~\cite{tmdmit1} that disorder effects in this system are weak. Neglecting the spinon Fermi surface contribution to the total resistivity, we approximate the full resistivity at criticality by $\Delta\rho_{xx}$. The factor of $N^2/2$ enhancement in $\Delta\rho_{xx}$ may offer a possible explanation for the large critical resistivity observed in Ref.~\cite{tmdmit1}.

Another key distinction from the Mott transition described by Eq.~\eqref{eq:_parton_vrtx_sc} lies in the scaling behavior of the bipartite spin fluctuations $\mathcal{F}^{s}$ in the Mott insulating phase. Due to the vortex condensation, both $\tilde{a}^{\uparrow}$ and $\tilde{a}^{\downarrow}$ are Higgsed, leading to a boundary-law scaling $\mathcal{F}^{s}\sim L$.

\section{Discussion and Outlook}
\label{sec:_Summary}

In this paper, we investigate the bipartite fluctuations of conserved quantities across interaction-driven ``metal-insulator transitions'' from a Landau FL to a ``charge insulator'' defined by localization criteria (see Sec.~\ref{subsec:_SSF_SWM}). These charge insulators, which exhibit a vanishing Drude weight, include incompressible examples like spin-liquid MIs and compressible ones like CFLs. If the transitions are continuous, we predict a logarithmic divergence of the localization tensor as the correlation length approaches infinity, as described in Eq.~\eqref{eq:_QM_log-div}. This divergence is associated with a universal subleading corner contribution to bipartite fluctuations (see Eq.~\eqref{eq:_BF_CFS}), with the universal coefficient determined by the critical resistivity jump, as outlined in Eq.~\eqref{eq:_C_jump}. Our results reveal a generalized fluctuation-response relation between ground-state charge (or spin) fluctuations and transport properties.

These findings stem from the general structure imposed by the Ioffe-Larkin composition rule and the presence of two dynamically decoupled parton sectors at criticality: one forms a stable Fermi-surface state, while the other behaves like a CFT. These results are expected to hold at other similar phase transitions as well.

Below, we provide further discussions on our central results and related topics while proposing interesting directions for future research.

\subsection{Non-Fermi Liquids}
\label{subsec:_Summary_NFL}

In Sec.~\ref{sec:_CFLFL} and Sec.~\ref{sec:_Mott}, we computed the static structure factors at the critical points (e.g., Eq.~\eqref{eq:_CFLFL_SSF} and Eq.~\eqref{eq:_Mott_SSF}) within the RPA framework. We expect these results to remain valid as long as gauge fluctuations are not too strong. The Ioffe-Larkin rule $\Pi^{-1}=\Pi_{f}^{-1}+\Pi_{b}^{-1}$, which arises from local gauge constraints, is expected to hold at the level of linear response~\cite{lee_nagaosa}. Regarding the expression for $\Pi_{f}$, prior studies~\cite{nayaknfl1,nayaknfl2,nfl_response_1994,AIMnfl,nfl3,nfl10} have shown that the charge susceptibility of the Fermi surface does not exhibit any singular contributions. We therefore anticipate that the long-wavelength behavior of $\Pi^{\tau\tau}$ remains dominated by the bosonic contribution $\Pi_{b}$ in the CFT sector, whose structure is rigidly constrained and given by Eq.~\eqref{eq:_CFT_response_1}. To further support this conclusion, one could attempt to compute the full expression for $\Pi_{f}$ at finite $\omega$ and $\boldsymbol{k}$ by including higher-order diagrams, as partially done in Refs.~\cite{nfl_response_1994,LU1_2,nflSYK2}. However, existing calculations are typically restricted to the $\boldsymbol{k}\rightarrow0$ limit relevant for conductivity. Extending these computations beyond this limit is technically demanding, and we leave such analysis to future work.

We have not considered the transitions between Landau FLs and charge-density-wave insulators, each coexisting with a neutral Fermi surface~\cite{frac_mit,cflfl3}. In such cases, the coupling of the density-wave order parameter $O$ to the Fermi surface of $f$-fermions leads to a different Landau damping term $\int_{k}|\omega||O(\omega,\mathbf{k})|^{2}$, which becomes relevant if the scaling dimension satisfies $\Delta[O]<1$. This scenario is likely true, and there is no dynamical decoupling of the boson and fermion sectors at criticality. It is possible that the Landau damping eventually drives the transition to become weakly first-order. Alternatively, there could be a new fixed point for the charge sector with a dynamical exponent $z>1$, as explored in a technically similar context in Ref.~\cite{FS_DW}. At this stage, we lack the theoretical tools to definitively determine the nature of these transitions, so the study of bipartite fluctuations across these exotic quantum critical points will need to be addressed in future research.

While this paper focuses on critical Fermi surfaces in two spatial dimensions, continuous Mott transitions between Fermi-liquid metals and spin-liquid insulators in three dimensions can also be constructed using similar theoretical frameworks. For example, Ref.~\cite{cmit3d} employs the same parton construction as described in Eq.~\eqref{eq:_parton2}, leading to a critical theory analogous to the two-dimensional case. In three dimensions, the interaction between spinons and chargons is found to be marginally irrelevant, and the chargon condensation falls into the 4D XY universality class. Since the system reaches its upper critical dimension, the scaling forms of various physical quantities exhibit logarithmic corrections. Within the vortex theory framework, the U(1) disorder operator is represented by a Wilson surface. When examining its shape dependence, additional singular geometries arise, including corners, cones, and trihedral vertices. It would be interesting to investigate whether universal subleading terms also appear in the bipartite fluctuations of critical Fermi surfaces in 3+1D systems.

In Sec.~\ref{sec:_vrtx}, we present a unified parton framework for critical Fermi surfaces by introducing vortices in Landau FLs. A conceptually equivalent but technically distinct approach exists for constructing the vortex theory without using partons. To illustrate this idea, consider the CFL-FL transition. One may begin with the non-linear bosonization of spinless FLs \cite{nlbosonization} (also see App.~\ref{App:_NL_Bosonization} for a brief review). Vortex excitations are then incorporated through the theory
\begin{flalign}
\mathcal{L}=\mathcal{L}_{\textrm{FS}}[\phi]+(a+A)\cdot J[\phi]-\frac{\mathtt{i}}{2\pi}\tilde{a}\wedge\textrm{d}a+\mathcal{L}_{\textrm{vrtx}}[\psi,\tilde{a}].
\label{eq:_bosonization_vrtx}
\end{flalign}
Here, $\mathcal{L}_{\textrm{FS}}[\phi]$ is the nonlinear theory introduced in Eq.~\eqref{eq:_U(1)_FS_action}, and $J[\phi]$ represents the U(1) current. The dynamical gauge field $a_{\mu}$ appears as a Lagrange multiplier that enforces the duality relation $J[\phi]=\frac{\mathtt{i}}{2\pi}\star\textrm{d}\tilde{a}$. By considering the vortex sector as in Eq.~\eqref{eq:_vrtx_f_1}, one can again realize the CFL-FL transition. The key insight is that the bosonic field $\phi$ encompasses a large number of low-energy excitations, including modes represented by the partons. A notable advantage of the formulation in Eq.~\eqref{eq:_bosonization_vrtx} lies in its capacity for systematic generalization by modifying $J[\phi]$, for example to incorporate higher-angular-momentum channels, as mentioned in Ref.~\cite{mit_dual}.

\subsection{Quantum Hall Physics}
\label{subsec:_Summary_QH}

In Sec.~\ref{sec:_CFLFL}, we examine the CFL phase, where the localization tensor takes the isotropic form $\mathsf{G}^{ij}=\mathsf{G}\delta^{ij}$ and a subleading corner contribution is identified in Eq.~\eqref{eq:_BF_QH_2}. Based on the standard RPA analysis of the HLR theory, we find that the value of $\mathsf{G}$ is universal and is given by the Hall conductivity $\sigma_{xy}$ (see Eq.~\eqref{eq:_QM_Hall}). However, as we will clarify in an upcoming work~\cite{Corner_Geom}, the situation is different for general (anisotropic) charge insulators. In such cases, $\mathsf{G}^{ij}$ becomes non-universal and is constrained by both a topological lower bound and an energetic upper bound~\cite{SWM2000,Resta2002Rev,Resta2011Rev,Fu2024-1} (see also App.~\ref{App:_Optical_Bounds})
\begin{flalign}
0&\leq\mathsf{G}^{ij}+\frac{\mathtt{i}}{2}\sigma_{xy}\epsilon^{ij},\nonumber\\0&\leq\lim_{\boldsymbol{A}\rightarrow0}\frac{1}{2\Delta}\frac{1}{V}\left\langle \!\frac{\partial^{2}H}{\partial A_{i}\partial A_{j}}\!\right\rangle -\mathsf{G}^{ij},
\label{eq:_universal_bounds}
\end{flalign}
where the right-hand side of each line is a positive semi-definite matrix. Here, $\Delta$ denotes the optical gap of the many-body Hamiltonian $H$, $V$ is the volume of the system, and $\boldsymbol{A}$ represents the background U(1) field. The specific conditions under which these bounds are saturated will be discussed further in Ref.~\cite{Corner_Geom}.

Under certain circumstances, the corner contribution to bipartite fluctuations can serve as a diagnostic for QH transitions. For instance, the expression $\mathsf{G}^{ij}=\delta^{ij}|\sigma_{xy}|/2$ is guaranteed by Galilean invariance in systems governed by microscopic Hamiltonians of the form~\cite{Fu2024-3,Corner_Geom}
\begin{flalign}
H=\sum_{\mathsf{j}}\frac{\boldsymbol{\pi}_{\mathsf{j}}^{2}}{2m}+\sum_{\mathsf{i}\neq\mathsf{j}}U(|\boldsymbol{r}_{\mathsf{i}}-\boldsymbol{r}_{\mathsf{j}}|),
\label{eq:_Kohn_Hamiltonian}
\end{flalign}
where $\boldsymbol{\pi}_{\mathsf{j}}=\boldsymbol{p}_{\mathsf{j}}-\boldsymbol{A}(\boldsymbol{r}_{\mathsf{j}})$ is the kinetic momentum of $\mathsf{j}$-th particle with mass $m$, and $U$ is an arbitrary interaction potential depending only on pairwise distances. Any QH state realized by such a UV description satisfies the bipartite fluctuations described by Eq.~\eqref{eq:_BF_QH_2} and  Eq.~\eqref{eq:_QM_Hall}. Consequently, changes in the corner term can signal topological phase transitions. In real materials, the bounds in Eq.~\eqref{eq:_universal_bounds} are typically not saturated. However, if one starts with a gapped QH state and varies the parameters, a sudden drop in the value of $\mathsf{G}^{ij}$ below the topological lower bound would still signal a topological transition. Such a transition—between a fractional Chern insulator and a charge-density-wave state driven by a displacement field—was recently discussed in Ref.~\cite{Fu2024-4} in the context of twisted bilayer semiconductors.

Although in this paper, we mainly discuss transitions at half-filling, the critical theories for the CFL-FL transition at other filling factors can be easily formulated within the vortex theory framework described in Sec.~\ref{sec:_vrtx}. Specifically, the case of $\nu=1/4$ can be described by Eq.~\eqref{eq:_parton_vrtx} together with the vortex sector 
\begin{flalign}
\mathcal{L}_{\textrm{vrtx}}[\psi,\tilde{a}]=\sum_{I=1}^{4}\bar{\psi}_{I}\slashed{\textrm{D}}_{\tilde{a}}\psi_{I}+\frac{1}{2e^{2}}\textrm{d}\tilde{a}\wedge\star\textrm{d}\tilde{a}.
\label{eq:_vrtx_f_2}
\end{flalign}
Through the fermionic particle-vortex duality~\cite{son2015, dual_review}, this theory can be demonstrated to be dual to the critical theory proposed in Ref.~\cite{cflfl3} based on a parton construction, where the chargon is fractionalized into four fermions. At the critical point, we again observe the bipartite charge fluctuations described by Eq.~\eqref{eq:_BF_CFS}, which exhibit a universal corner contribution related to the critical resistivity jumps via Eq.~\eqref{eq:_C_jump}.

\subsection{Entanglement Properties}
\label{subsec:_Summary_EE}

The study of the corner contribution to the entanglement entropy in CFTs has a longer history~\cite{CFTEE1,CFTEE2,CFTEE3,CFTEE4}. Considering the shape shown in \figref{fig:_corner}, the entanglement entropy takes the form $\#L-\textrm{s}(\theta)\log(L)+\ldots$ where $\#$ is again non-universal. While the analytical expression for $\textrm{s}(\theta)$ is not known in general, it has been shown that, in the limit where $\theta$ approaches $\pi$, $\textrm{s}(\theta)$ is proportional to $C_{T}(\pi - \theta)^{2}$, where $C_{T}$ is the stress-tensor central charge of the CFT~\cite{CFTEE3,CFTEE4}. The universal subleading term has been utilized in Monte Carlo studies of deconfined quantum critical points~\cite{DQCPEE1,DQCPEE2,DQCPEE3,Mengdisop5}, as well as symmetric mass generation~\cite{Mengdisop4}. However, there are subtleties in correctly extracting the subleading term and accounting for contributions from Goldstone modes~\cite{ExtCorEE1,ExtCorEE2}.

The entanglement entropy in the CFL phase at $\nu=1/2$ has been numerically investigated in Ref.~\cite{CFLEE1,CFLEE2}. Both of their results exhibit the $L\log(L)$ scaling, resembling that of free fermions, although their overall coefficients differ by a factor of two. This problem was recently revisited by Ref.~\cite{CFLEE3}, which provided evidence supporting an enhancement in entanglement compared to free fermions. This result clearly deviates from our result of a boundary law $\mathcal{F}^{c}\sim L$ for bipartite charge fluctuations in Sec.~\ref{subsubsec:_BF_CFL}. A similar situation arises in the gapless Mott insulator, where the entanglement entropy has been numerically observed to exhibit a logarithmic enhancement~\cite{FS-SL_EE}. As we have seen in Sec.~\ref{sec:_Mott}, although the charge fluctuations scale as $\mathcal{F}^{c}\sim L$, the spinon Fermi surface still gives rise to the spin fluctuations $\mathcal{F}^{s}\sim L\log(L)$. Therefore, it would be interesting to investigate the bipartite fluctuations of other quantities in the CFL phase that potentially identify the scaling of its entanglement entropy.

The entanglement entropy in both phases separated by the critical points discussed in Sec.~\ref{sec:_CFLFL} and Sec.~\ref{sec:_Mott} scales as  $L\log(L)$. It is therefore natural to expect the critical points to follow the same scaling, although further careful technical studies are needed. While critical Fermi surfaces and CFTs share similarities in their bipartite charge fluctuations, they are likely to be distinguished by their entanglement entropy.

In the context of non-interacting fermions, the full entanglement spectrum is entirely encoded in the bipartite fluctuations and the higher-order cumulants\footnote{More precisely, the full entanglement spectrum of non-interacting fermions depends only on cumulants with even $n$~\cite{EEBF1,EEBF2,EEBF3,EEBF4,EEBF5,EEBF6,EEBF7,EEBF8,EEBF9,EEvsBF}.}
\begin{flalign}
\mathcal{N}_{\mathcal{A}}^{[n]}=\lim_{\vartheta\rightarrow0}(-\mathtt{i}\partial_{\vartheta})^{n}\log\langle\mathcal{W}_{\mathcal{A}}(\vartheta)\rangle,
\label{eq:_cumulant}
\end{flalign}
where $\mathcal{W}_{\mathcal{A}}(\vartheta)$ denotes the U(1) disorder operator
\begin{flalign}
\mathcal{W}_{\mathcal{A}}(\vartheta)=\exp\left(\vartheta\int_{\mathcal{A}}\star J\right).
\label{eq:_dis_op}
\end{flalign}
In the dual vortex theory, the disorder operator corresponds to the Wilson loop Eq.~\eqref{eq:_WL}, where $\star J=\frac{\mathtt{i}}{2\pi}\textrm{d}\tilde{a}$. A comprehensive understanding of the higher-order cumulants in critical Fermi surfaces requires a systematic extension of the Ioffe-Larkin rule to gauge-invariant non-linear response functions. It would be interesting to explore whether there are any distinctive features between even and odd values of $n$, as observed in Ref.~\cite{FCS_Corner} for integer QH and Laughlin states.

A closely related topic is the role of the density three-point function in probing Fermi-surface topology in free fermion systems~\cite{Tam2022,Tam2024}. It would be interesting to explore how this result generalizes to the non-Fermi liquids considered in this work. Moreover, this may offer a new perspective for distinguishing critical Fermi surfaces from CFTs, since $\langle JJJ \rangle$ vanishes identically for an abelian conserved spin-1 current $J$ in any 3D CFT~\cite{JJJ_CFT3_2013}.

\subsection{Numerics and Experiments}
\label{subsec:_Summary_EX_Num}

One of our key findings highlights the change in scaling behaviors of bipartite fluctuations (or U(1) disorder operators) across interaction-driven ``metal-insulator transitions,'' which closely mirrors transitions between different phases of higher-form symmetries. Despite recent progress in understanding certain aspects of the low-energy physics of Fermi surfaces from the perspective of LU(1) anomaly~\cite{LU1_1}, and efforts to formulate Fermi-surface dynamics using nonlinear bosonization via the infinite-dimensional Lie group of canonical transformations~\cite{nlbosonization} (also see App.~\ref{App:_NL_Bosonization}), a generalized symmetry principle for the quantum phase transitions considered in this paper remains an open question. Nevertheless, our quantitative results in Eq.~\eqref{eq:_QM_log-div} and Eq.~\eqref{eq:_BF_CFS} have direct implications for numerical and experimental studies of unconventional quantum criticalities in metals.

Our analytical result Eq.~\eqref{eq:_BF_CFS}, which identifies a universal corner contribution with a universal constant $C_{\rho}$, provides a numerical method (via the Monte Carlo approach) to self-consistently verify whether a proposed candidate microscopic model realizes the desired critical field theory. In the context of CFTs, this method has been benchmarked using the Bose-Hubbard model~\cite{chengdisop} and applied to study unconventional phase transitions, such as the symmetric mass generation transition of Dirac fermions~\cite{Mengdisop4}. The numerical method based on our approach would be especially valuable for verifying the realizability of intriguing quantum critical points in metals, where powerful techniques such as the conformal bootstrap~\cite{bootstrapRMP,bootstrapRMP2} are not applicable.

Significant progress has recently been made in the search for microscopic realizations of the CFL-FL transition. On the theoretical side, a candidate model based on spatially modulated Landau levels was discussed in Ref.~\cite{cfl_drude}, where tuning the interaction strength leads to a transition characterized by the vanishing of the Drude weight. Although whether the transition is continuous remains to be determined, this setup provides a promising platform for exploring critical behaviors. In parallel, experimental signatures consistent with zero-field CFL phases and CFL-FL transitions have been observed in twisted $\textrm{MoTe}_{2}$ bilayers~\cite{CFLFL_Xu1,CFLFL_Xu2} and multilayer graphene systems~\cite{CFLFL_Ju}. These developments offer new opportunities for both numerical and experimental tests of the universal scaling behaviors discussed in this work.

Several experimental techniques have been proposed for measuring bipartite fluctuations. In cold atoms and quantum gases, correlation functions and bipartite fluctuations can be directly probed using high-resolution imaging techniques (see e.g.~\cite{BFEX1,BFEX2,BFEX3,BFEX4,BFEX5}). For 1+1D systems exhibiting space-time duality~\cite{EEBF2,EEBF9}, charge fluctuations can be obtained through quantum point contact measurements~\cite{QPC_EX1,QPC_EX2}.

In quantum materials, the localization tensor $\mathsf{G}$ (i.e., the many-body quantum metric), which governs the corner term in $2+1$ dimensions~\cite{Corner_Geom,Corner_Geom2}, can also be experimentally accessed. For example, $\mathsf{G}$ can be extracted from structure factor measurements using inelastic X-ray scattering, as demonstrated by recent experimental data for the insulator LiF~\cite{QMEX1}. Furthermore, it can be obtained through frequency sum rules from optical responses, with a detailed analysis for the $\textrm{MnBi}_{2}\textrm{Te}_{4}$ film provided in Ref.~\cite{QM_MnBi2Te4}. These techniques suggest feasible pathways for experimentally testing the universal predictions and investigating the broader landscape of unconventional quantum criticalities discussed in this work.


\section*{Acknowledgment}

We thank Luca Delacrétaz, Dominic Else, Hart Goldman, Chao-Ming Jian, Prashant Kumar, Michael Levin, Zlatko Papić, Dam Thanh Son, and Senthil Todadri for related discussions. We thank Cenke Xu for discussions and participation in the early stage of the project, and Meng Cheng for communicating unpublished results. This work was supported in part by the Simons Collaboration on Ultra-Quantum Matter, which is a grant from the Simons Foundation (651442), and the Simons Investigator award (990660).

{\it Note added:} {\it (1)} We would like to draw the reader’s attention to a related paper~\cite{disop_FS} that appeared in the same arXiv listing as the first version of this manuscript. {\it (2)} We would like to mention a subsequent work~\cite{CFLEE3} that numerically verifies the predictions for CFLs (Eq.~\eqref{eq:_BF_QH_2} along with Eq.~\eqref{eq:_QM_Hall}) through wavefunction-based Monte Carlo simulations. {\it (3)} We thank Kang-Le Cai, Meng Cheng, and Prashant Kumar for a related collaboration~\cite{Corner_Geom}, which addresses the corner term in charge insulators and its relation to many-body quantum geometry.

\begin{widetext}

\appendix

\section{$\textrm{LU}(1)$ Anomaly and Structure Factor}
\label{App:_LU1}

In this appendix, we provide a geometric interpretation of the static structure factor, drawing from the $\textrm{LU}(1)$ anomaly of Fermi-surface states~\cite{LU1_1,LU1_2}. We begin by introducing the low-energy patch theory
\begin{flalign}
\mathcal{L}=\int\textrm{d}\theta\psi^{\dagger}(\tau,\boldsymbol{x},\theta)(\textrm{D}_{\tau}+\mathtt{i}v_{F}^{j}(\theta)\textrm{D}_{j}+\kappa_{ij}(\theta)\textrm{D}_{i}\textrm{D}_{j})\psi(\tau,\boldsymbol{x},\theta)+(\textrm{interactions}).
\end{flalign}
Here, $\theta$ denotes an angle variable labeling the patch, $v_{F}^{j}(\theta)$ represents the fermi velocity, and $\kappa_{ij}(\theta)$ is the curvature tensor. The gauge covariant derivative is denoted by $\textrm{D}_{\mu}=\partial_{\mu}-\mathtt{i}A_{\mu}$, where $A_{\mu}$ is the background electromagnetic field. Under the scaling limit, the paramagnetic current (at $A=0$) is given by
\begin{flalign}
J^{\tau}(\tau,\boldsymbol{x})&=\frac{\delta\mathcal{S}}{\delta A_{\tau}}=-\mathtt{i}\int\textrm{d}\theta\rho(\tau,\boldsymbol{x},\theta),\nonumber\\J^{i}(\tau,\boldsymbol{x})&=\frac{\delta\mathcal{S}}{\delta A_{i}}=\int\textrm{d}\theta\rho(\tau,\boldsymbol{x},\theta)v_{F}^{i}(\theta), \label{eq:_LU1_current}
\end{flalign}
where $\rho(\tau,\boldsymbol{x},\theta)=\psi^{\dagger}(\tau,\boldsymbol{x},\theta)\psi(\tau,\boldsymbol{x},\theta)$ represents the density at each patch of the Fermi surface. We introduce the phase-space current density $\mathcal{J}^{\mu}$ such that $J^{\mu}(\tau,\boldsymbol{x})=\int\textrm{d}\theta\mathcal{J}^{\mu}(\tau,\boldsymbol{x},\theta)$. Due to the $\textrm{LU}(1)$ anomaly~\cite{LU1_1}, the Ward identity (in Euclidean signature) is as follows
\begin{flalign}
\partial_{\mathtt{I}}\mathcal{J}^{\mathtt{I}}=\frac{-\mathtt{i}}{8\pi^{2}}\varepsilon^{\mathtt{IJKL}}\partial_{\mathtt{I}}\mathcal{A}_{\mathtt{J}}\partial_{\mathtt{K}}\mathcal{A}_{\mathtt{L}},
\label{eq:_LU1_Ward_id}
\end{flalign}
where $\partial_{\mathtt{I}}=(\partial_{\tau},\partial_{\boldsymbol{x}},\partial_{\theta})$. As for the phase-space background field $\mathcal{A}_{\mathtt{I}}$, we introduce 
\begin{flalign}
\mathcal{A}_{\tau}=A_{\tau}(\tau,\boldsymbol{x}),\qquad\mathcal{A}_{\boldsymbol{x}}=A_{\boldsymbol{x}}(\tau,\boldsymbol{x})+\boldsymbol{k}_{F}(\theta),\qquad\mathcal{A}_{\theta}\textrm{ is independent of }\tau,\boldsymbol{x}
\end{flalign}
Here, $A_{\mu}=(A_{\tau},A_{\boldsymbol{x}})$ is the ordinary background electromagnetic field. We provide two explanations for the inclusion of $\boldsymbol{k}_{F}(\theta)$ in $\mathcal{A}_{\boldsymbol{x}}$. {\it (1)} We want to turn on a background flux $F_{\boldsymbol{x}\boldsymbol{k}}$ that ensures the canonical commutation relation $[x_{i},k_{j}]=\mathtt{i}\delta_{ij}$~\cite{LU1_You}. {\it (2)} It aligns with the semiclassical equation of motion $\dot{\boldsymbol{k}}=E$. (One may also check Sec. VI. B of Ref.~\cite{LU1_1}). The LU(1) Ward identity Eq. \eqref{eq:_LU1_Ward_id} leads to 
\begin{equation}
\partial_{\tau}\mathcal{J}^{\tau}(\tau,\boldsymbol{x},\theta)+\partial_{\boldsymbol{x}}\cdot\mathcal{J}^{\boldsymbol{x}}(\tau,\boldsymbol{x},\theta)=\frac{-\mathtt{i}}{(2\pi)^{2}}\boldsymbol{E}\times\frac{\textrm{d}\boldsymbol{k}_{F}(\theta)}{\textrm{d}\theta},
\end{equation}
where $\boldsymbol{E}=-F_{\tau\boldsymbol{x}}=\partial_{\boldsymbol{x}}A_{\tau}-\partial_{\tau}A_{\boldsymbol{x}}$.
We have used $\partial_{\theta}J^{\theta}=0$ in the presence of a
background electric field \citep{LU1_1}. We focus
on the response to $A_{\tau}$ and set $A_{\boldsymbol{x}}=0$. In
the momentum space, this is given by 
\begin{equation}
(-\mathtt{i}\omega+\boldsymbol{v}_{F}\cdot\boldsymbol{k})\rho(\omega,\boldsymbol{k},\theta)=\frac{-\mathtt{i}}{(2\pi)^{2}}\left(\boldsymbol{k}\times\frac{\textrm{d}\boldsymbol{k}_{F}(\theta)}{\textrm{d}\theta}\right)A_{\tau}(\omega,\boldsymbol{k}).
\end{equation}
The equal-time response of the total electric charge density $J^{\tau}(\omega,\boldsymbol{k})=-\mathtt{i}\int\textrm{d}\theta\rho(\omega,\boldsymbol{k},\theta)$ is therefore 
\begin{equation}
\Pi^{\tau\tau}(\tau\rightarrow0,\boldsymbol{k})=\frac{1}{(2\pi)^{2}}\int\frac{\textrm{d}\omega}{2\pi}\int_{0}^{2\pi}\textrm{d}\theta\frac{-1}{\mathtt{i}\omega-\boldsymbol{v}_{F}(\theta)\cdot\boldsymbol{k}}\left(\boldsymbol{k}\times\frac{\textrm{d}\boldsymbol{k}_{F}(\theta)}{\textrm{d}\theta}\right)=\frac{\textrm{Area}}{(2\pi)^{2}},
\label{eq:_LU1_Pi00_1}
\end{equation}
where $(\textrm{Area})$ represents the value of the shaded area in \figref{fig:_ssf_geom}. The dependence on the fermi velocity $\boldsymbol{v}_{F}$ is eliminated after the $\omega$-integral, and $\boldsymbol{k}\times\partial_{\theta}\boldsymbol{k}_{F}(\theta)$ has the geometric interpretation of an area element.

For a spherical Fermi surface $\boldsymbol{k}_{F}(\theta)=k_{F}(\cos\theta,\sin\theta)$,
we have $\boldsymbol{k}\times\partial_{\theta}\boldsymbol{k}_{F}(\theta)=\boldsymbol{k}\cdot\boldsymbol{k}_{F}(\theta)$,
and accordingly 
\begin{equation}
\Pi^{\tau\tau}(\tau\rightarrow0,\boldsymbol{k})=\frac{1}{(2\pi)^{2}}\int\frac{\textrm{d}\omega}{2\pi}\int_{0}^{2\pi}\textrm{d}\theta\frac{-k_{F}|\boldsymbol{k}|\cos\theta}{\mathtt{i}\omega-v_{F}|\boldsymbol{k}|\cos\theta}=\frac{k_{F}|\boldsymbol{k}|}{2\pi}\int_{0}^{2\pi}\frac{\textrm{d}\theta}{2\pi}\frac{|\cos\theta|}{2}=\frac{k_{F}|\boldsymbol{k}|}{2\pi^{2}}.
\label{eq:_LU1_Pi00_2}
\end{equation}
This exactly reproduces the free-fermion result in Eq.~\eqref{eq:_FS_Pi00}. It's worth noting that one can introduce interactions that explicitly break the LU(1) symmetry, such as scattering between fermion modes from different patches. For instance, considering the RPA treatment of a local density-density interaction, we obtain the density response 
\begin{equation}
\hat{\Pi}^{\tau\tau}(\mathtt{i}\omega,\boldsymbol{k})=\frac{\Pi^{\tau\tau}(\mathtt{i}\omega,\boldsymbol{k})}{1+u\Pi^{\tau\tau}(\mathtt{i}\omega,\boldsymbol{k})},
\end{equation}
Here, $\Pi^{\tau\tau}(\mathtt{i}\omega,\boldsymbol{k})$ represents the free-fermion result from Eq.~\eqref{eq:_FS_response}, and $u$ denotes a coupling constant. Consequently, the equal-time correlation function becomes
\begin{equation}
\int\frac{\textrm{d}\omega}{2\pi}\hat{\Pi}^{\tau\tau}(\mathtt{i}\omega,\boldsymbol{k})=\frac{k_{F}|\boldsymbol{k}|}{2\pi^{2}}\frac{\sqrt{2\tilde{u}+1}+2\tilde{u}\arctan(\sqrt{2\tilde{u}+1})}{(2\tilde{u}+1)^{3/2}},
\end{equation}
where $\tilde{u}=u\mathscr{D}_{F}$ is rescaled by the density of states $\mathscr{D}_{F}=\frac{k_{F}}{2\pi v_{F}}$ at the fermi level. Also, note that in the problem of the half-filled Landau level, the Ward identity Eq.~\eqref{eq:_LU1_Ward_id} needs modification due to the presence of an emergent gauge field~\cite{LU1_1}. Therefore, the geometric interpretation presented here should not conflict with the results in Sec.~\ref{subsubsec:_BF_CFL}.

\section{Response Theory and Coulomb Gauge}
\label{App:_Coulomb}


The imaginary‑time (Matsubara) response function of a $\mathrm{U}(1)$-conserved current is defined as 
\begin{equation}
\Pi^{\mu\nu}(x-y)=\left.-\frac{\delta^{2}\log\mathcal{Z}[A]}{\delta A_{\mu}(x)\delta A_{\nu}(y)}\right|_{A=0}=\left.\left\langle \frac{\delta^{2}\mathcal{S}[A]}{\delta A_{\mu}(x)\delta A_{\nu}(y)}\right\rangle \right|_{A=0}-\langle J^{\mu}(x)J^{\nu}(y)\rangle_{c}
\end{equation}
where $\mathcal{Z}[A]=\int\mathcal{D}[\textrm{matter}]e^{-\mathcal{S}[\textrm{matter},A]}$ is the partition function in the presence of the background field $A_\mu$, and 
\begin{equation}
\langle J^{\mu}(x)J^{\nu}(y)\rangle_{c}=\langle J^{\mu}(x)J^{\nu}(y)\rangle-\langle J^{\mu}(x)\rangle\langle J^{\nu}(y)\rangle
\end{equation}
denotes the connected two-point function of the paramagnetic current operator, which itself is defined as
\begin{equation}
J^{\mu}(x)=\left.\frac{\delta\mathcal{S}[A]}{\delta A_{\mu}(x)}\right|_{A=0}.
\end{equation}

For nonrelativistic electrons, the diamagnetic contact term is absent in the density channel in both the Landau FL and insulating phases, i.e., $\delta^{2}\mathcal{S}/(\delta A_{\tau}\delta A_{\tau})=0$.  By applying the Källén-Lehmann spectral representation, one can show that the static structure factor defined in the real-time formalism coincides with the equal-time limit $\tau\rightarrow0^{+}$ of the Matsubara density response function.

For any translationally and rotationally invariant systems in 2+1 dimensions, the Fourier transform of the response function, i.e., $\Pi^{\mu\nu}(k)=\int\textrm{d}^{3}xe^{-\mathtt{i}x\cdot k}\Pi^{\mu\nu}(x)$, has the structure
\begin{flalign}
\Pi^{\tau\tau}(k)&=\frac{|\boldsymbol{k}|^{2}}{\omega^{2}}\Pi_{L}(\omega,|\boldsymbol{k}|),\nonumber\\\Pi^{\tau i}(k)&=-\frac{k_{i}}{\omega}\Pi_{L}(\omega,|\boldsymbol{k}|)+\varepsilon^{ij}k_{j}\mathsf{H}(\omega,|\boldsymbol{k}|),\nonumber\\\Pi^{ij}(k)&=\frac{k_{i}k_{j}}{|\boldsymbol{k}|^{2}}\Pi_{L}(\omega,|\boldsymbol{k}|)+\left(\delta^{ij}-\frac{k_{i}k_{j}}{|\boldsymbol{k}|^{2}}\right)\Pi_{T}(\omega,|\boldsymbol{k}|)+\omega\varepsilon^{ij}\mathsf{H}(\omega,|\boldsymbol{k}|),\label{eq:_response_2d}
\end{flalign}
where $\Pi_{L}$ and $\Pi_{T}$ describe the longitudinal and transverse components respectively, while $\mathsf{H}$ characterizes the Hall response. Here, $k=(\omega,\boldsymbol{k})$ collectively denotes the Matsubara frequency and momentum. The $\mathrm{U}(1)$ Ward identity $k_{\mu}\Pi^{\mu\nu}(k)=0$ is automatically satisfied.

To simplify the analysis, it is useful to decompose the spatial components of
the background/gauge field into longitudinal and transverse components, expressed
as $\boldsymbol{A}=\boldsymbol{A}_{L}+\boldsymbol{A}_{T}$ where 
\begin{flalign}
\boldsymbol{A}^{L} & (k)=\mathcal{P}^{L}\boldsymbol{A}(k),\qquad\mathcal{P}_{ij}^{L}=\frac{k_{i}k_{j}}{|\boldsymbol{k}|^{2}},\nonumber \\
\boldsymbol{A}^{T} & (k)=\mathcal{P}^{T}\boldsymbol{A}(k),\qquad\mathcal{P}_{ij}^{T}=\delta_{ij}-\mathcal{P}_{ij}^{L}.
\end{flalign}
It is convenient to introduce a scalar field $A_{T}$ to represent the transverse component $A_{i}^{T}$, defined by 
\begin{equation}
A_{i}^{T}(k)=\frac{\varepsilon_{ij}k_{j}}{|\boldsymbol{k}|}A_{T}(k)\quad\textrm{or}\quad A_{T}(k)=\frac{\varepsilon_{ij}k_{j}}{|\boldsymbol{k}|}A_{i}^{T}(k).
\end{equation}
In the Coulomb gauge, where $\boldsymbol{A}^{L} = 0$, the response theory takes a simpler form when expressed in the $(A_{\tau}, A_{T})$ basis
\begin{flalign}
 \mathcal{S}[A]=\int\frac{\textrm{d}^{3}k}{(2\pi)^{3}}\frac{1}{2}(A_{\tau}(-k)\;A_{T}(-k))\left(\begin{array}{cc}
\Pi^{\tau\tau}(k) & \mathsf{H}(k)|\boldsymbol{k}|\\
\mathsf{H}(k)|\boldsymbol{k}| & \Pi^{TT}(k)
\end{array}\right)\left(\begin{array}{c}
A_{\tau}(k)\\
A_{T}(k)
\end{array}\right)\label{eq:_response_2d-Coulomb}
\end{flalign}
with $\Pi^{\tau\tau}(k)=\frac{|\boldsymbol{k}|^{2}}{\omega^{2}}\Pi_{L}(\omega,|\boldsymbol{k}|)$ and 
$\Pi^{TT}(k)=-\Pi_{T}(\omega,|\boldsymbol{k}|)$. Notice that the Chern-Simons term is 
\begin{flalign}
\int\frac{-\mathtt{i}}{4\pi}A\wedge\textrm{d}A=\int\frac{\textrm{d}^{3}k}{(2\pi)^{3}}\frac{1}{2}(A_{\tau}(-k)\;A_{T}(-k))\frac{-|\boldsymbol{k}|}{2\pi}\left(\begin{array}{cc}
0 & 1\\
1 & 0
\end{array}\right)\left(\begin{array}{c}
A_{\tau}(k)\\
A_{T}(k)
\end{array}\right).
\end{flalign}

\section{Ioffe-Larkin Composition Rule} \label{App:_Ioffe-Larkin}

We investigate the response function of electrons using the parton theory described in Eq. \eqref{eq:_parton_theory}.
In the standard RPA approach, integrating out both fermionic and bosonic
partons $f$ and $b$ leads to the effective action 
\begin{equation}
\mathcal{S}=\int_{k}\frac{\Pi_{f}^{\mu\nu}(k)}{2}(a_{\mu}(-k)+e_{f}A_{\mu}(-k))(a_{\nu}(k)+e_{f}A_{\nu}(k))+\frac{\Pi_{b}^{\mu\nu}(k)}{2}(-a_{\mu}(-k)+e_{b}A_{\mu}(-k))(-a_{\nu}(k)+e_{b}A_{\nu}(k)).
\end{equation}
To enforce gauge invariance, we integrate out the dynamical gauge
fields $a_{\mu}$ at the RPA level. This yields the total response
theory $\mathcal{L}=\int_{k}\frac{1}{2}\Pi^{\mu\nu}(k)A_{\mu}(-k)A_{\nu}(k)$
with
\begin{flalign}
\Pi & =e_{f}^{2}\Pi_{f}+e_{b}^{2}\Pi_{b}-(e_{f}\Pi_{f}-e_{b}\Pi_{b})(\Pi_{f}+\Pi_{b})^{-1}(e_{f}\Pi_{f}-e_{b}\Pi_{b})\nonumber \\
 & =e_{f}^{2}(\Pi_{f}-\Pi_{f}(\Pi_{f}+\Pi_{b})^{-1}\Pi_{f})+e_{b}^{2}(\Pi_{b}-(\Pi_{f}+\Pi_{b})^{-1}\Pi_{b})+e_{b}e_{f}(\Pi_{f}(\Pi_{f}+\Pi_{b})^{-1}\Pi_{b}+\Pi_{b}(\Pi_{f}+\Pi_{b})^{-1}\Pi_{f})\nonumber \\
 & =(e_{f}+e_{b})^{2}\Pi_{f}(\Pi_{f}+\Pi_{b})^{-1}\Pi_{b}=\Pi_{f}(\Pi_{f}+\Pi_{b})^{-1}\Pi_{b},
\end{flalign}
where we have used $e_{f}+e_{b}=1$. Therefore, we find the celebrated Ioffe-Larkin
composition rule~\cite{Ioffe-Larkin}
\begin{equation}
\Pi^{-1}=\Pi_{f}^{-1}+\Pi_{b}^{-1},\label{eq:_Ioffe-Larkin_fb}
\end{equation}
which is independent of the specific assignment $(e_{f},e_{b})$ of the global
$\textrm{U}(1)$ charge among partons. 

In the Coulomb gauge, the parton response functions $\Pi_{f}$ and $\Pi_{b}$ can be written
as 
\begin{equation}
\Pi_{f}=\left(\begin{array}{cc}
\Pi_{f}^{\tau\tau}(k) & \mathsf{H}_{f}|\boldsymbol{k}|\\
\mathsf{H}_{f}|\boldsymbol{k}| & \Pi_{f}^{TT}(k)
\end{array}\right),\qquad\Pi_{b}=\left(\begin{array}{cc}
\Pi_{b}^{\tau\tau}(k) & \mathsf{H}_{b}|\boldsymbol{k}|\\
\mathsf{H}_{b}|\boldsymbol{k}| & \Pi_{b}^{TT}(k)
\end{array}\right).
\end{equation}
Applying the Ioffe-Larkin rule Eq.~\eqref{eq:_Ioffe-Larkin_fb}, the resulting gauge-invariant response functions are
\begin{flalign}
\Pi^{\tau\tau} & =\frac{\Pi_{f}^{\tau\tau}\det(\Pi_{b})+\Pi_{b}^{\tau\tau}\det(\Pi_{f})}{\det(\Pi_{f}+\Pi_{b})},\nonumber \\
\Pi^{TT} & =\frac{\Pi_{f}^{TT}\det(\Pi_{b})+\Pi_{b}^{TT}\det(\Pi_{f})}{\det(\Pi_{f}+\Pi_{b})},\nonumber \\
\Pi^{\tau T} & =\Pi^{T\tau}=|\boldsymbol{k}|\frac{\mathsf{H}_{f}\det(\Pi_{b})+\mathsf{H}_{b}\det(\Pi_{f})}{\det(\Pi_{f}+\Pi_{b})}.\label{eq:_Ioffe-Larkin_2}
\end{flalign}
In time-reversal invariant systems, where $\mathsf{H}_{f}=0$
and $\mathsf{H}_{b}=0$, these simplify further to
\begin{equation}
\Pi^{\tau\tau}=\frac{\Pi_{f}^{\tau\tau}\Pi_{b}^{\tau\tau}}{\Pi_{f}^{\tau\tau}+\Pi_{b}^{\tau\tau}},\qquad\Pi^{TT}=\frac{\Pi_{f}^{TT}\Pi_{b}^{TT}}{\Pi_{f}^{TT}+\Pi_{b}^{TT}},\qquad\Pi^{\tau T}=\Pi^{T\tau}=0.\label{eq:_Ioffe-Larkin_3}
\end{equation}

\section{Scaling of Bipartite Fluctuations}
\label{app:_BF_scaling}


In this appendix, we present the technical details of gauge-invariant regularization schemes used to handle the UV divergence in evaluating Eq.~\eqref{eq:_BF_WL}. Our starting point is the static structure factor of electrons. At long wavelengths, we assume rotational invariance and consider a generic power-law correlation Eq.~\eqref{eq:_power-law_Pi00} in real space. To keep our discussion as general as possible, we allow the value of $\alpha$ to be continuously tuned with $\alpha>2$. 


After the Fourier transformation of Eq.~\eqref{eq:_power-law_Pi00}, the static structure factor is given by
\begin{flalign}
\int\textrm{d}^{d}x\frac{-C_{0}}{|\boldsymbol{x}|^{\alpha}}e^{\mathtt{i}\boldsymbol{k}\cdot\boldsymbol{x}}=\frac{2^{d-\alpha}\pi^{\frac{d}{2}}\Gamma(\frac{d-\alpha}{2})}{\Gamma(\frac{\alpha}{2})}\frac{-C_{0}}{\left|\boldsymbol{k}\right|^{d-\alpha}}=\begin{cases}
-2\pi C_{0}\log(1/|\boldsymbol{k}|) & \alpha=2\\
+2\pi C_{0}|\boldsymbol{k}| & \alpha=3\\
+\frac{\pi}{2}C_{0}|\boldsymbol{k}|^{2}\log(1/|\boldsymbol{k}|) & \alpha=4\\
-\frac{2\pi}{9}C_{0}|\boldsymbol{k}|^{3} & \alpha=5\\
-\frac{\pi}{32}C_{0}|\boldsymbol{k}|^{4}\log(1/|\boldsymbol{k}|) & \alpha=6\\
\qquad\vdots & \quad\vdots
\end{cases}
\label{eq:_power-law_SSF}
\end{flalign}
where $d=2$. Although Eq.~\eqref{eq:_power-law_SSF} includes the Fourier relation for the case $\alpha=2$, physical systems with a global U(1) symmetry must satisfy $\alpha>2$, since the static structure is required to vanish in the limit $\boldsymbol{k}\rightarrow0$. This constraint can be understood by noting that, in this limit, the density operator $J^{\tau}(\tau,\boldsymbol{k}\rightarrow0)$ corresponds to the globally conserved charge $Q=\mathtt{i}\int\textrm{d}^{2}\boldsymbol{x}J^{\tau}(\tau,\boldsymbol{x})$. As a result, the equal-time density-density correlation is simply the charge variance
\begin{flalign}
\textrm{Var}(Q)=\langle Q^{2}\rangle-\langle Q\rangle^{2}
\end{flalign}
evaluated in the zero-temperature ground state. Since $[Q,H]=0$, the ground state of the Hamiltonian $H$ must also be an eigenstate of $Q$, implying that $\textrm{Var}(Q)=0$.


In the case of gapless modes in CFLs, the static structure factor contains a contribution of the form $|\boldsymbol{k}|^{3}\log(1/|\boldsymbol{k}|)$, which leads to an additional term $|\boldsymbol{x}|^{-5}\log|\boldsymbol{x}|$ in real space, alongside Eq.~\eqref{eq:_power-law_Pi00} with $\alpha=5$. However, this logarithmic correction does not change the scaling of bipartite fluctuations. This issue is further discussed in App.~\ref{subapp:_BF_scaling_k}.


\subsection{Real-Space Method}
\label{subapp:_BF_scaling_x}

In App.~\ref{subapp:_BF_scaling_x}, we propose a real-space method that is suitable for both dimensional regularization and UV cut-off schemes. The idea is to embed the equal-time correlation function Eq.~\eqref{eq:_power-law_Pi00} into a space-time current-current correlation
\begin{flalign}
\langle\check{J}^{\mu}(x)\check{J}^{\nu}(0)\rangle=\frac{C_{0}}{|x|^{\alpha}}\left(\delta^{\mu\nu}-\frac{\alpha}{\alpha-2}\frac{x^{\mu}x^{\nu}}{|x|^{2}}\right),
\label{eq:_JJ_reg}
\end{flalign}
such that $\Pi^{\tau\tau}(\tau\rightarrow0,\boldsymbol{x})=-\langle\check{J}^{\tau}(\tau,\boldsymbol{x})\check{J}^{\tau}(\tau,0)\rangle$. We once again introduce a dual gauge field $\check{a}_{\mu}$ to represent the current $\check{J}^{\mu}=\frac{\mathtt{i}}{2\pi}\varepsilon^{\mu\nu\rho}\partial_{\nu}\check{a}_{\rho}$. The gauge-field propagator $\check{D}_{\mu\nu}(x-y)=\langle\check{a}_{\mu}(x)\check{a}_{\nu}(y)\rangle$ can be written as 
\begin{flalign}
\check{D}_{\mu\nu}^{(\alpha)}(x)&=\frac{(2\pi)^{2}C_{0}}{(\alpha-2)^{2}}\frac{1}{|x|^{\alpha-2}}\left((1+\zeta)\delta^{\mu\nu}-\zeta(\alpha-2)\frac{x^{\mu}x^{\nu}}{|x|^{2}}\right)\nonumber\\&=\frac{(2\pi)^{2}C_{0}}{(\alpha-2)^{2}}\left(\frac{\delta^{\mu\nu}}{|x|^{\alpha-2}}+\zeta\partial_{\mu}\partial_{\nu}\begin{cases}
\frac{1}{4-\alpha}|x|^{4-\alpha} & \alpha\neq4\\
\log|x| & \alpha=4
\end{cases}\right),
\label{eq:_aa_reg}
\end{flalign}
where $\zeta$ is a Faddeev-Popov gauge-fixing parameter. One can replace the gauge field $\tilde{a}_{\mu}$ by $\check{a}_{\mu}$ in the Wilson loop Eq.~\eqref{eq:_WL}, and calculate the bipartite fluctuations Eq.~\eqref{eq:_BF_WL} by 
\begin{flalign}
\mathcal{F}_{\mathcal{A}}^{(\alpha,\epsilon)}=\frac{1}{(2\pi)^{2}}\int_{\mathcal{C}}\textrm{d}x^{i}\int_{\mathcal{C}}\textrm{d}y^{j}\check{D}_{ij}^{(\alpha)}(\epsilon,\boldsymbol{x}-\boldsymbol{y}),
\label{eq:_BF_WL_reg}
\end{flalign}
where a small splitting $\epsilon>0$ in the ``temporal direction'' serves as a small real-space UV cut-off, and the integrals are performed along the closed spatial loop $\mathcal{C}=\partial\mathcal{A}$. In Eq.~\eqref{eq:_BF_WL_reg}, there are two parameters, $\epsilon$ and $\alpha$. If the cut-off $\epsilon$ is strictly set to zero, and the power $\alpha$ matches the physical value from the Ioffe-Larkin rule Eq.~\eqref{eq:_Ioffe-Larkin_fv}, the generalized formula Eq.~\eqref{eq:_BF_WL_reg} exactly reduces back to the original definition Eq.~\eqref{eq:_BF_WL}.

Now, we are prepared to address the gauge-invariant calculation of Eq.~\eqref{eq:_BF_WL_reg}. In the UV cut-off scheme, we fix the value of $\alpha$ and treat $\epsilon$ as a small expansion parameter. Let us examine the square geometry of the loop $\mathcal{C}$, which has a side length of $L$. The case of CFTs where $\alpha=4$ has been extensively discussed in Ref.~\cite{wulog}. It has a leading cut-off-dependent boundary-law term together with a universal subleading logarithmic term
\begin{flalign}
\frac{\mathcal{F}_{\mathcal{A}}^{(4,\epsilon)}}{C_{0}}=\frac{\pi}{4}\frac{|\mathcal{C}|}{\epsilon}-2\log|\mathcal{C}|+\textrm{const},
\end{flalign}
where $|\mathcal{C}|=4L$ is the perimeter of the square. By setting $\alpha=3$ and following the calculations in Eq. (9)-(12) from Ref.~\cite{wulog}, we obtain the result for Landau FLs
\begin{flalign}
\frac{\mathcal{F}_{\mathcal{A}}^{(3,\epsilon)}}{C_{0}}=2|\mathcal{C}|(\log|\mathcal{C}|+\textrm{const}),
\end{flalign}
As a self-consistency check, the final result of $\mathcal{F}_{\mathcal{A}}$ is again independent of the Faddeev-Popov gauge-fixing parameter $\zeta$. We have also evaluated Eq.~\eqref{eq:_BF_WL_reg} for generic values of $\alpha$ other than 3 and 4
\begin{flalign}
\frac{\mathcal{F}_{\mathcal{A}}^{(\alpha,\epsilon)}}{C_{0}}=|\mathcal{C}|\epsilon^{3-\alpha}\frac{\sqrt{\pi}\Gamma(\frac{\alpha-3}{2})}{(\alpha-2)^{2}\Gamma(\frac{\alpha-2}{2})}+|\mathcal{C}|^{4-\alpha}\frac{(2^{\frac{\alpha}{2}}-2-2(\alpha-4)\textrm{Hypergeometric2F1}(1,\frac{3-\alpha}{2};\frac{3}{2};-1))}{2^{5-\frac{3\alpha}{2}}\Gamma(\alpha-1)/\Gamma(\alpha-4)}.
\label{eq:_BF_square_gen}
\end{flalign}
The general expression Eq.~\eqref{eq:_BF_square_gen} is informative in understanding the contribution of gapless modes in different phases of matter. In non-local systems where the instantaneous charge correlation decays even slower than Landau FLs (i.e., $2<\alpha<3$), the first boundary-law term in Eq.~\eqref{eq:_BF_square_gen} vanishes as $\epsilon$ approaches zero, leaving only the second term, which is independent of $\epsilon$ and scales as $|\mathcal{C}|^{4-\alpha}$. Conversely, when $\alpha>3$, the leading-order term is always given by the boundary law $|\mathcal{C}|\epsilon^{3-\alpha}$, which contains a power-law UV divergence. Another notable observation is that when the instantaneous charge correlation is weaker than in CFTs (i.e., $\alpha>4$), the universal subheading term $|\mathcal{C}|^{4-\alpha}$ vanishes in the large-$|\mathcal{C}|$ limit. As we will see, this holds true for CFLs. In \tabref{tab:_BF_scaling}, we summarize our findings under the cut-off regularization scheme.

If one is only interested in the ``universal term'' that remains independent of any UV cut-off, there is another convenient regularization scheme that is in the same spirit as dimensional regularization. Here, we set $\epsilon=0$ and retain $\alpha$ as an arbitrary parameter in the integrals for Eq.~\eqref{eq:_BF_WL_reg}. Subsequently, we consider the final result through an expansion in terms of small $\delta=\alpha-\bar{\alpha}$, where $\bar{\alpha}$ represents the physical value. In this scheme, all power-law UV divergences are automatically eliminated, and the logarithmic divergence manifests as $\delta^{-1}$. Let us check two simple geometries, a square and a circle. One can easily find the result for a square
\begin{flalign}
\frac{\mathcal{F}_{\mathcal{A}}^{(\alpha,0)}}{C_{0}}&=(\textrm{the 2nd term in Eq.~\eqref{eq:_BF_square_gen}})=\begin{cases}
2|\mathcal{C}|(\log|\mathcal{C}|-\delta^{-1}-2\log(2)+\sqrt{2}-\sinh^{-1}(1)) & \alpha=3\\
2(\log|\mathcal{C}|-\delta^{-1})+\log(32)-\frac{\pi}{2}-4 & \alpha=4\\
|\mathcal{C}|^{-1}(\frac{48-32\sqrt{2}}{9})\rightarrow0 & \alpha=5
\end{cases},
\end{flalign}
and the result for a circle
\begin{flalign}
\frac{\mathcal{F}_{\mathcal{A}}^{(\alpha,0)}}{C_{0}}&=|\mathcal{C}|^{4-\mathit{\alpha}}\frac{\pi^{\mathit{\alpha}-\frac{5}{2}}\Gamma(\frac{3}{2}-\frac{\mathit{\alpha}}{2})}{2(\alpha-2)\Gamma(3-\frac{\alpha}{2})}=\begin{cases}
2|\mathcal{C}|(\log|\mathcal{C}|-\delta^{-1}-\log(\frac{\pi}{2})) & \alpha=3\\
-\frac{\pi^{2}}{2} & \alpha=4\\
-\frac{\pi^{2}}{3}|\mathcal{C}|^{-1}(\log|\mathcal{C}|-\delta^{-1}+\frac{1}{6}(5-\log(\frac{\pi^{6}}{64})))\rightarrow0 & \alpha=5
\end{cases},
\end{flalign}
where $|\mathcal{C}|$ represents the perimeter of $\mathcal{C}=\partial\mathcal{A}$ in the both cases. There are some lessons we can learn from this exercise. {\it (1)} When $\alpha=3$, the coefficient of the leading term $|\mathcal{C}|\log|\mathcal{C}|$ remains independent of detailed geometries. {\it (2)} Both the subleading term $|\mathcal{C}|$ in the case of $\alpha=3$ and the universal constant term in the case of $\alpha=4$ depend on detailed geometries. {\it (3)} When $\alpha=5$, the universal term depends on the geometry but always vanishes under the large-$|\mathcal{C}|$ scaling. {\it (4)} When $\alpha=4$, the term $\log|\mathcal{C}|$ only appears when the geometry contains sharp corners. 

In fact, the corner contribution can also be conveniently calculated using this regularization scheme, compared to the calculation under the cut-off scheme in Ref.~\cite{wulog}. It is convenient to choose the gauge $\zeta=\frac{1}{\alpha-3}$, ensuring that contribution from the same straight line vanishes. Considering the correlation between two straight lines, we find that the contribution exhibits a logarithmic divergence only when $\alpha=3$ and $\alpha=4$. For $\alpha=4$, the angle dependence Eq.~\eqref{eq:_angle_function} can also be exactly reproduced.

\subsection{Momentum-Space Method}
\label{subapp:_BF_scaling_k}

Starting with the propagator $\tilde{D}_{\mu\nu}=\langle\tilde{a}_{\mu}\tilde{a}_{\nu}\rangle$ for the gauge field $\tilde{a}_{\mu}$ as defined in Eq.~\eqref{eq:_dual_current}, the bipartite fluctuations described in Eq.~\eqref{eq:_BF_WL} can be rewritten as
\begin{flalign}
\mathcal{F}_{\mathcal{A}}=\frac{1}{(2\pi)^{2}}\int\frac{\textrm{d}\omega\textrm{d}^{2}\boldsymbol{k}}{(2\pi)^{3}}Y_{\mathcal{C}}^{i}(-\boldsymbol{k})D_{ij}(\mathtt{i}\omega,\boldsymbol{k})Y_{\mathcal{C}}^{j}(\boldsymbol{k})
\label{eq:_BF_WL_2}
\end{flalign}
where $Y_{\mathcal{C}}^{i}(\boldsymbol{k})$ is a linear functional over the closed loop $\mathcal{C}=\partial\mathcal{A}$, defined by
\begin{flalign}
Y_{\mathcal{C}}^{i}(\boldsymbol{k})=\int_{\mathcal{C}}\textrm{d}x^{i}e^{\mathtt{i}\boldsymbol{k}\cdot\boldsymbol{x}}.
\end{flalign}
Taking advantage of the divergence-free property $k_{i}Y_{\mathcal{C}}^{i}(\boldsymbol{k})=-\mathtt{i}\int_{\mathcal{C}}\textrm{d}x^{i}\partial_{i}e^{\mathtt{i}\boldsymbol{k}\cdot\boldsymbol{x}}=0$, we can write $Y_{\mathcal{C}}^{i}=\varepsilon^{ij}k_{j}\overline{Y}_{\mathcal{C}}$, where 
\begin{flalign}
\overline{Y}_{\mathcal{C}}(\boldsymbol{k})=\mathtt{i}\int_{\mathcal{A}}\textrm{d}^{2}\boldsymbol{x}e^{\mathtt{i}\boldsymbol{k}\cdot\boldsymbol{x}}.
\end{flalign}
Consequently, Eq.~\eqref{eq:_BF_WL_2} can be equivalently expressed as
\begin{flalign}
\mathcal{F}_{\mathcal{A}}=\int\frac{\textrm{d}^{2}\boldsymbol{k}}{(2\pi)^{2}}\Pi^{\tau\tau}(\tau\rightarrow0,\boldsymbol{k})|\overline{Y}_{\mathcal{C}}(\boldsymbol{k})|^{2},
\label{eq:_BF_WL_3}
\end{flalign}
where $\Pi^{\tau\tau}(\tau\rightarrow0,\boldsymbol{k})$ is the static structure factor. 

We are particularly interested in the term $\left|\boldsymbol{k}\right|^{\beta}$ or $\left|\boldsymbol{k}\right|^{\beta}\log|\boldsymbol{k}|$ that appears in the static structure factor. We find that the Fourier transform of $\left|\boldsymbol{k}\right|^{\beta}$ is nonzero only when $\beta$ is an odd integer
\begin{flalign}
\int\frac{\textrm{d}^{d}\boldsymbol{k}}{(2\pi)^{d}}e^{-\mathtt{i}\boldsymbol{k}\cdot\boldsymbol{x}}\left|\boldsymbol{k}\right|^{\beta}=\frac{2^{\beta}\pi^{\frac{1-d}{2}}\Gamma(\frac{d+\beta}{2})}{\Gamma(\frac{d-1}{2})\Gamma(1-\frac{d+\beta}{2})}\frac{1}{|\boldsymbol{x}|^{d+\beta}}=\begin{cases}
0 & \beta=0\\
-\frac{1}{2\pi}|\boldsymbol{x}|^{-3} & \beta=1\\
0 & \beta=2\\
+\frac{9}{2\pi}|\boldsymbol{x}|^{-5} & \beta=3\\
0 & \beta=4\\
\qquad\vdots & \quad\vdots
\end{cases}
\label{eq:_power-law_SSF_2}
\end{flalign}
where $d=2$. As for $\left|\boldsymbol{k}\right|^{\beta}\log|\boldsymbol{k}|$, we find nonzero results for all integer values of $\beta$
\begin{flalign}
\int\frac{\textrm{d}^{d}\boldsymbol{k}}{(2\pi)^{d}}e^{-\mathtt{i}\boldsymbol{k}\cdot\boldsymbol{x}}\left|\boldsymbol{k}\right|^{\beta}\log|\boldsymbol{k}|&=\frac{2^{\beta}\pi^{\frac{1-d}{2}}\Gamma(\frac{d+\beta}{2})}{\Gamma(\frac{d-1}{2})\Gamma(1-\frac{d+\beta}{2})}\frac{\Psi^{(0)}(\frac{d+\beta}{2})+\Psi^{(0)}(1-\frac{d+\beta}{2})-2\log(|\boldsymbol{x}|/2)}{2|\boldsymbol{x}|^{d+\beta}}\nonumber\\&=\begin{cases}
-\frac{1}{2\pi|\boldsymbol{x}|^{2}} & \beta=0\\
+\frac{\log|\boldsymbol{x}|+\log2+\gamma_{E}-2}{2\pi|\boldsymbol{x}|^{3}} & \beta=1\\
+\frac{2}{\pi|\boldsymbol{x}|^{4}} & \beta=2\\
-\frac{3(3\log|\boldsymbol{x}|+\log8+3\gamma_{E}-8)}{2\pi|\boldsymbol{x}|^{5}} & \beta=3\\
-\frac{32}{\pi|\boldsymbol{x}|^{6}} & \beta=4\\
\qquad\vdots & \quad\vdots
\end{cases}
\label{eq:_power-law_SSF_3}
\end{flalign}
where again $d=2$, $\Psi^{(0)}(z)=\frac{\textrm{d}}{\textrm{d}z}\log\Gamma(z)$ denotes the the polygamma function, and $\gamma_{E}$ is the Euler-Mascheroni constant. The cases from Eq.~\eqref{eq:_power-law_SSF_2} with odd $\beta$ and those from Eq.~\eqref{eq:_power-law_SSF_3} with even $\beta$ together account for all the cases considered earlier in Eq.~\eqref{eq:_power-law_SSF}. The new cases arise in Eq.~\eqref{eq:_power-law_SSF_3} with odd $\beta$, where the power-law correlation receives a logarithmic correction.

We can evaluate Eq.~\eqref{eq:_BF_WL_3} using dimensional regularization, treating $\beta$ as a tunable parameter in the integral. To illustrate this, consider a circle $\mathcal{C}$ with a radius $R$. In this case, we have 
\begin{flalign}
\overline{Y}_{\mathcal{C}}(\boldsymbol{k})=\mathtt{i}\int_{0}^{R}r\textrm{d}r\int_{0}^{2\pi}\textrm{d}\theta e^{\mathtt{i}|\boldsymbol{k}|r\cos\theta}=\frac{\mathtt{i}2\pi R}{|\boldsymbol{k}|}\textrm{BesselJ}(1,|\boldsymbol{k}|R).
\end{flalign}
For the term proportional to $\left|\boldsymbol{k}\right|^{\beta}$, the contribution to to bipartite fluctuations is given by
\begin{flalign}
\mathcal{F}_{\mathcal{A}}\sim\int\frac{\textrm{d}^{2}\boldsymbol{k}}{(2\pi)^{2}}|\boldsymbol{k}|^{\beta}|\overline{Y}_{\mathcal{C}}(\boldsymbol{k})|^{2}=2\pi R^{2}\int_{0}^{+\infty}\textrm{d}kk^{\beta-1}\textrm{BesselJ}(1,|\boldsymbol{k}|R)^{2}=\frac{\sqrt{\pi}\Gamma(1+\frac{\beta}{2})\Gamma(\frac{1}{2}-\frac{\beta}{2})}{\Gamma(1-\frac{\beta}{2})\Gamma(2-\frac{\beta}{2})}R^{2-\beta}.
\end{flalign}
For the term proportional to $\left|\boldsymbol{k}\right|^{\beta}\log|\boldsymbol{k}|$, the contribution becomes
\begin{flalign}
\mathcal{F}_{\mathcal{A}}&\sim\int\frac{\textrm{d}^{2}\boldsymbol{k}}{(2\pi)^{2}}|\boldsymbol{k}|^{\beta}\log|\boldsymbol{k}||\overline{Y}_{\mathcal{C}}(\boldsymbol{k})|^{2}=2\pi R^{2}\int_{0}^{+\infty}\textrm{d}kk^{\beta-1}\log(k)\textrm{BesselJ}(1,|\boldsymbol{k}|R)^{2}\nonumber\\&=\frac{\sqrt{\pi}\Gamma(1+\frac{\beta}{2})\Gamma(\frac{1}{2}-\frac{\beta}{2})(\frac{2}{\beta-2}+\Psi^{(0)}(\frac{1}{2}-\frac{\beta}{2})-2\Psi^{(0)}(1-\frac{\beta}{2})-\Psi^{(0)}(1+\frac{\beta}{2})+2\log R)}{(\beta-2)\Gamma(1-\frac{\beta}{2})^{2}}R^{2-\beta}.
\end{flalign}
Applying these results to some examples of compressible states, we find the cut-off-independent contributions  
\begin{flalign}
\begin{cases}
|\boldsymbol{k}|\quad\Rightarrow\quad2R\log R & \textrm{Landau FLs and superfluids}\\
|\boldsymbol{k}|^{3}\log(1/|\boldsymbol{k}|)\quad\Rightarrow\quad\frac{3}{8}\frac{(\log R)^{2}}{R}\rightarrow0 & \textrm{gapless modes in CFLs}\\
|\boldsymbol{k}|\log(1/|\boldsymbol{k}|)\quad\Rightarrow\quad R(\log R)^{2} & \textrm{certain non-Fermi liquids}
\end{cases}.
\end{flalign}
Therefore, in CFLs, the subleading term from gapless modes vanishes under large-scale scaling. For the leading term, since $|\boldsymbol{x}|^{-4}<|\boldsymbol{x}|^{-5}\log|\boldsymbol{x}|<|\boldsymbol{x}|^{-5}$ at long distances, and both $|\boldsymbol{x}|^{-4}$ and $|\boldsymbol{x}|^{-5}$ lead to a boundary law, the logarithmic correction does not change the leading-order scaling. This conclusion has also been independently confirmed in Ref.~\cite{disop_FS} using a momentum-space cut-off scheme. The correction involving $|\boldsymbol{x}|^{-3}\log|\boldsymbol{x}|$ (or $|\boldsymbol{k}|\log(1/|\boldsymbol{k}|)$) deserves special attention, as $|\boldsymbol{x}|^{-3}$ (or $|\boldsymbol{k}|$) already exhibits a logarithmic violation of the boundary law. An example of such ``double-log violation'' of the boundary law, $R(\log R)^{2}$, was first pointed out in Ref.~\cite{disop_FS}.

\section{Optical Absorption and Structure Factor}
\label{App:_Optical_Bounds}

For gapped charge insulators, the topological lower bound in Eq.~\eqref{eq:_universal_bounds} can be understood from the positive semi-definiteness of the many-body quantum geometric tensor, where the quantum metric and Berry curvature correspond to its real and imaginary parts, respectively~\cite{SWM2000,Resta2002Rev,Resta2011Rev,Corner_Geom,Fu2024-2}. For gapless charge insulators (with a finite $\mathsf{G}^{ij}$), such as CFLs, we provide an additional argument based on the positivity of optical absorption power, directly following the discussion in Ref.~\cite{Fu2024-1,Fu2024-3} for gapped states.

We perturb the system with a monochromatic electric field $\boldsymbol{E}(t)=\mathcal{E}^{*}e^{\mathtt{i}\omega t}+\mathcal{E}e^{-\mathtt{i}\omega t}$.
The absorption power is 
\begin{equation}
P\sim\int\textrm{d}t\boldsymbol{j}(t)\cdot\boldsymbol{E}(t)=\mathcal{E}^{*}\cdot\boldsymbol{j}(\omega)+\mathcal{E}\cdot\boldsymbol{j}^{*}(\omega)=\mathcal{E}_{i}^{*}(\sigma^{ij}(\omega)+\sigma^{*ji}(\omega))\mathcal{E}_{j}.\label{eq:_absorp_power}
\end{equation}
It is convenient to define the absorptive part of the conductivity as
\begin{equation}
\sigma_{\textrm{abs}}^{ij}(\omega)=\frac{\sigma^{ij}(\omega)+\sigma^{*ji}(\omega)}{2}=\textrm{Re}\sigma_{+}^{ij}(\omega)+\mathtt{i}\textrm{Im}\sigma_{-}^{ij}(\omega),\label{eq:_absorp_conductivity}
\end{equation}
and the associated optical weight as
\begin{equation}
W_{[\ell]}^{ij}=\int_{0}^{+\infty}\frac{\textrm{d}\omega}{\pi}\frac{\sigma_{\textrm{abs}}^{ij}(\omega)}{\omega^{\ell}}.\label{eq:_absorp_weight}
\end{equation}
Here, we use $\sigma_{\pm}^{ij}=(\sigma^{ij}\pm\sigma^{ji})/2$ to
denote the longitudinal and Hall conductivities, respectively. Assuming the unperturbed system is in the ground state, the absorption
power given by Eq. \ref{eq:_absorp_power} is expected to be non-negative,
i.e., $P\sim\mathcal{E}_{i}^{*}\sigma_{\textrm{abs}}^{ij}(\omega)\mathcal{E}_{j}\geq0$
holds for any complex vector $\mathcal{E}$. Consequently, both $\sigma_{\textrm{abs}}^{ij}(\omega)$
at each frequency and the optical weight $W_{[\ell]}^{ij}$, as given
in Eq. \ref{eq:_absorp_weight}, are positive semi-definite matrices.

The real part of $W_{[1]}^{ij}$ is given by the SWM sum rule in Eq.~\eqref{eq:_SWM_sum}, while the imaginary part of $W_{[1]}^{ij}$ is related to the DC Hall
conductivity $\sigma^{xy}=\textrm{Re}\sigma_{-}^{xy}(0)$ through the Kramers-Kronig relation
\begin{equation}
\textrm{Re}\sigma_{-}^{ij}(\omega)=\frac{2}{\pi}\int_{0}^{+\infty}\textrm{d}\nu\frac{\nu}{\nu^{2}-\omega^{2}}\textrm{Im}\sigma_{-}^{ij}(\nu).
\end{equation}
In conclusion, one obtains the positive semi-definite matrix 
\begin{equation}
0\leq W_{[1]}^{ij}=\mathsf{G}^{ij}+\frac{\mathtt{i}}{2}\textrm{Re}\sigma_{-}^{ij}(0),
\end{equation}
which corresponds to the lower bound stated in Eq. \ref{eq:_universal_bounds}. In two dimensions, using $\det(W_{[1]})\geq0$, one can show that
\begin{equation}
\frac{\textrm{Tr}(\mathsf{G})}{2}\geq\sqrt{\det(\mathsf{G})}\geq\frac{|\sigma^{xy}|}{2}.
\end{equation}

For gapped charge insulators, $W_{[1]}^{ab}$ is precisely the many-body
quantum geometric tensor, as defined by flux insertion or twisted
boundary conditions. Importantly, the lower bound is expected to hold as long as the absorption power in Eq.~\ref{eq:_absorp_power} remains non-negative, even if the system is gapless and the DC Hall conductivity $\sigma_{xy}$ is not a topological invariant. For both gapless and gapped QH states, Ref.~\cite{Corner_Geom} clarifies the conditions for bound saturation from two perspectives: {\it (1)} invariance under continuous translations and Galilean boosts, and {\it (2)} the holomorphic properties of many-body wavefunctions, including Laughlin wavefunctions and composite-fermion Fermi sea wavefunctions.

\section{Dual Theories for CFL-FL Transition}
\label{App:_cflfl_dual}

We begin by establishing the duality relation between the vortex theory Eq.~\eqref{eq:_parton_vrtx} (together with Eq.~\eqref{eq:_vrtx_f_1}) and the critical theory Eq.~\eqref{eq:_cflfl_theory_1}. It is known that a single Dirac fermion enjoys the fermion-fermion duality~\cite{dual_review,son2015}
\begin{equation}
\bar{\psi}\slashed{\textrm{D}}_{A}\psi\quad\longleftrightarrow\quad\bar{\chi}\slashed{\textrm{D}}_{a}\chi-\frac{\mathtt{i}}{2\pi}a\textrm{d}b+\frac{\mathtt{i}2}{4\pi}b\textrm{d}b-\frac{\mathtt{i}}{2\pi}b\textrm{d}A+\frac{\mathtt{i}}{4\pi}A\textrm{d}A,
\end{equation}
where each Dirac fermion is defined through the Pauli-Villars scheme with another heavy Dirac fermion in
the UV. Upon integrating out the gauge field $b$, the resulting expression takes on the usual form found in the literature, albeit with incorrectly quantized topological terms
\begin{equation}
\bar{\psi}\slashed{\textrm{D}}_{A}\psi-\frac{1}{2}\frac{\mathtt{i}}{4\pi}A\textrm{d}A\quad\longleftrightarrow\quad\bar{\chi}\slashed{\textrm{D}}_{a}\chi-\frac{1}{2}\frac{\mathtt{i}}{4\pi}a\textrm{d}a-\frac{1}{2}\frac{\mathtt{i}}{2\pi}a\textrm{d}A.\label{eq:_ff_duality_1}
\end{equation}
Its time-reversal image yields yet another fermionic particle-vortex duality 
\begin{equation}
\bar{\psi}\slashed{\textrm{D}}_{A}\psi-\frac{1}{2}\frac{\mathtt{i}}{4\pi}A\textrm{d}A\quad\longleftrightarrow\quad\bar{\chi}\slashed{\textrm{D}}_{a}\chi-\frac{1}{2}\frac{\mathtt{i}}{4\pi}a\textrm{d}a+\frac{1}{2}\frac{\mathtt{i}}{2\pi}a\textrm{d}A.\label{eq:_ff_duality_2}
\end{equation}
Let us apply the duality Eq. \eqref{eq:_ff_duality_2} to the two fermions
in Eq.~\eqref{eq:_vrtx_f_1}, subject to the constraint $\psi_{1}^{\dagger}\psi_{1}=\psi_{2}^{\dagger}\psi_{2}$
\begin{flalign}
\bar{\chi}_{1}\slashed{\textrm{D}}_{\tilde{b}}\chi_{1}+\bar{\chi}_{2}\slashed{\textrm{D}}_{\tilde{c}}\chi_{2}-\frac{1}{2}\frac{\mathtt{i}}{4\pi}\tilde{b}\textrm{d}\tilde{b}-\frac{1}{2}\frac{\mathtt{i}}{4\pi}\tilde{c}\textrm{d}\tilde{c}+\frac{\mathtt{i}}{2\pi}\tilde{a}\textrm{d}(A-a+\tilde{b}/2+\tilde{c}/2)+\frac{\mathtt{i}}{4\pi}\tilde{a}\textrm{d}\tilde{a}+\frac{\mathtt{i}}{2\pi}\lambda\textrm{d}(\tilde{b}-\tilde{c}),
\end{flalign}
where the fluxes of $\tilde{b}$ and $\tilde{c}$ represent the densities of $\psi_{1}$ and $\psi_{2}$, and $\lambda$ serves as a Lagrangian multiplier. After integrating
out both $\lambda$ and $\tilde{a}$, we find that (when $\psi_{1}^{\dagger}\psi_{1}=\psi_{2}^{\dagger}\psi_{2}$)
\begin{equation}
\sum_{I=1}^{2}\bar{\psi}_{I}\slashed{\textrm{D}}_{\tilde{a}}\psi_{I}+\frac{\mathtt{i}}{2\pi}\tilde{a}\textrm{d}(A-a)\quad\longleftrightarrow\quad\sum_{I=1}^{2}\bar{\chi}_{I}\slashed{\textrm{D}}_{\tilde{b}}\chi_{I}-\frac{2\mathtt{i}}{4\pi}\tilde{b}\textrm{d}\tilde{b}-\frac{\mathtt{i}}{2\pi}\tilde{b}\textrm{d}(A-a)-\frac{\mathtt{i}}{4\pi}(A-a)\textrm{d}(A-a).
\label{eq:_ff_duality_3}
\end{equation}
Together with the Fermi-surface sector described by $\mathcal{L}_{\textrm{FS}}[f,a]$, we find Eq.~\eqref{eq:_parton_vrtx} (together with Eq.~\eqref{eq:_vrtx_f_1}) and Eq.~\eqref{eq:_cflfl_theory_1} are indeed related by the fermionic particle-vortex duality~\cite{dual_review,son2015}.

In view of the abelian duality web~\cite{dual_review}, there are other formulations of the critical theory as well. For a single Dirac fermion, there is a fermion-boson particle-vortex duality
\begin{equation}
\bar{\psi}\slashed{\textrm{D}}_{A}\psi\quad\longleftrightarrow\quad|\textrm{D}_{a}\varphi|^{2}+|\varphi|^{4}-\frac{\mathtt{i}}{4\pi}a\textrm{d}a-\frac{\mathtt{i}}{2\pi}a\textrm{d}A.\label{eq:_fb_duality_1}
\end{equation}
Using Eq. \eqref{eq:_fb_duality_1}, we can express the dual theory of $\bar{\psi}_{1}\slashed{\textrm{D}}_{\tilde{a}}\psi_{1}+\bar{\psi}_{2}\slashed{\textrm{D}}_{\tilde{a}}\psi_{2}-\frac{\mathtt{i}}{2\pi}\tilde{a}\textrm{d}a$ as follows
\begin{flalign}
 |\textrm{D}_{\tilde{b}}\varphi_{1}|^{2}+|\varphi_{1}|^{4}+|\textrm{D}_{\tilde{c}}\varphi_{2}|^{2}+|\varphi_{2}|^{4}-\frac{\mathtt{i}}{4\pi}\tilde{b}\textrm{d}\tilde{b}-\frac{\mathtt{i}}{4\pi}\tilde{c}\textrm{d}\tilde{c}-\frac{\mathtt{i}}{2\pi}\tilde{a}\textrm{d}(a+\tilde{b}+\tilde{c}).
\end{flalign}
Integrating out $\tilde{a}$ imposes the constraint $\tilde{c}=-a-\tilde{b}$, leading to the bosonic dual theory for the CFL-FL transition
\begin{flalign}
\mathcal{L}=\mathcal{L}_{\textrm{FS}}[f,a+A]+|\textrm{D}_{\tilde{b}}\varphi_{1}|^{2}+|\varphi_{1}|^{4}+|\textrm{D}_{-a-\tilde{b}}\varphi_{2}|^{2}+|\varphi_{2}|^{4}-\frac{2\mathtt{i}}{4\pi}\tilde{b}\textrm{d}\tilde{b}-\frac{\mathtt{i}}{2\pi}\tilde{b}\textrm{d}a-\frac{\mathtt{i}}{4\pi}a\textrm{d}a.
\end{flalign}
The phase transition is driven by the simultaneous condensation of $\varphi_{1}$ and $\varphi_{2}$. 

In addition, several other dual critical theories based on level-rank dualities have been reviewed in Ref.~\cite{cflfl3}.

\section{Static Structure Factor at Criticality}
\label{App:_CFS_SSF}

In this appendix, we provide technical details on evaluating the integrals for the static structure factors at the quantum critical points, as given by Eq.~\eqref{eq:_CFLFL_SSF} and Eq.~\eqref{eq:_Mott_SSF}. If one naively expands the integrand at small $\boldsymbol{k}$ before performing the integration, the leading-order term can be analytically integrated, yielding
\begin{flalign}
\Pi^{\tau\tau}(\tau\rightarrow0,\boldsymbol{k})\approx\int_{0}^{+\infty}\frac{\textrm{d}\omega}{\pi}\frac{|\boldsymbol{k}|^{2}k_{F}v_{F}(k_{F}v_{F}\sigma_{xx}^{b}+4\pi\omega(\sigma_{xx}^{b})^{2}-4\pi\omega(\sigma_{xy}^{b})^{2})}{\omega(k_{F}v_{F}+4\pi\omega\sigma_{xx}^{b}){}^{2}-16\pi^{2}\omega^{3}(\sigma_{xy}^{b})^{2}}=\frac{\sigma_{xx}^{b}}{\pi}|\boldsymbol{k}|^{2}\log(1/|\boldsymbol{k}|)+\ldots
\label{eq:_CFLFL_Pi00_v2}
\end{flalign}
where the logarithmic divergence arises from the $\omega$-integral. The factor $\log(1/|\boldsymbol{k}|)$ reflects the identical scaling of $\omega$ and $\boldsymbol{k}$ at the critical point. The analytic evaluation of Eq.~\eqref{eq:_Mott_SSF} proceeds similarly and can be viewed as a special case of Eq.~\eqref{eq:_CFLFL_Pi00_v2} with $\sigma_{xy}^{b}=0$.

\begin{figure}
\includegraphics[width=0.88\textwidth]{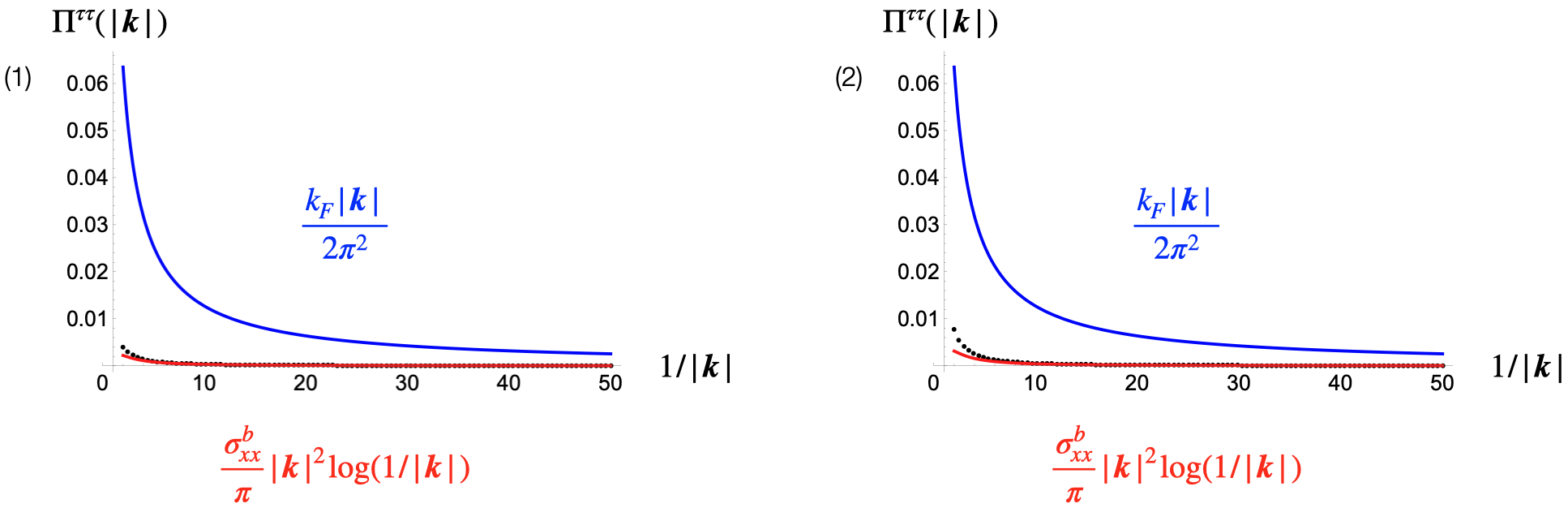}
\caption{{\it (1)} Numerical result (black) for the static structure factor at the CFL-FL transition, based on Eq.~\eqref{eq:_CFLFL_SSF}, with parameters $k_{F}=\sqrt{2\pi}$, $v_{F}=0.25$, and $\sigma_{xx}^{b}=-\sigma_{xy}^{b}=1/(8\pi)$; {\it (2)} Numerical result (black) for the static structure factor at the Mott transition, based on Eq.~\eqref{eq:_Mott_SSF}, with $k_{F}=\sqrt{2\pi}$, $v_{F}=0.25$, and $\sigma_{xx}^{b}=0.355155/(2\pi)$. In both panels, the results are compared with the CFT-like scaling behavior (red) and that of ordinary Fermi surfaces (blue).} \label{fig:_ssf_CFS}
\end{figure}

To verify the analytic result in Eq.~\eqref{eq:_CFLFL_Pi00_v2}, we also numerically evaluated the integrals using the exact expressions in Eq.~\eqref{eq:_CFLFL_SSF} and Eq.~\eqref{eq:_Mott_SSF}, with the results presented in \figref{fig:_ssf_CFS}. In these numerical calculations, the Fermi momentum $k_{F}$ is determined by the Luttinger theorem $V_{\textrm{FS}}/(2\pi)^{2}=1/2$ at half-filling\footnote{The Luttinger theorem, as it applies to the composite Fermi liquid and the spinon Fermi surface, is discussed in Ref.~\cite{LU1_1}. The more precise expressions are $V_{\textrm{FS}}/(2\pi)^{2}=(1/2)/(2\pi\ell_{B}^{2})$ and $V_{\textrm{FS}}/(2\pi)^{2}=(1/2)/V_{\textrm{cell}}$, where $\ell_{B}$ represents the magnetic length and $V_{\textrm{cell}}$ denotes the volume of the unit cell. Our purpose is to numerically verify the analytic result in Eq.~\eqref{eq:_CFLFL_Pi00_v2}. To this end, we set $2\pi\ell_{B}^{2} = 1$ and $V_{\textrm{cell}} = 1$ for simplicity.}, where the Fermi-surface volume is given by $V_{\textrm{FS}}=\pi k_{F}^{2}$. We choose a small $v_{F}$ to facilitate the numerical estimation. For the chargon conductivity $\sigma_{xx}^{b}$ at the Mott transition, we adopt the critical conductivity value obtained from the conformal bootstrap for the 3D XY universality class~\cite{bootstrap}. For chargons at the CFL-FL transition, following Ref.~\cite{cflfl3}, we crudely estimate that the universal transverse and Hall resistivities $\rho_{xx}^{b},\rho_{xy}^{b}$ are both of order $4\pi$. In both cases, the numerical results agree well with the analytic predictions at small $\boldsymbol{k}$, confirming that the static structure factor at critical Fermi surfaces exhibits CFT-like scaling and deviates from that of ordinary Fermi surfaces.

\section{Non-Linear Bosonization} \label{App:_NL_Bosonization}

In this appendix, we offer a brief introduction to non-linear bosonization~\cite{nlbosonization} of Fermi liquids from the perspective of coherent-state construction. For any fermionic systems with translation symmetry, we can introduce the fermion bilinear operator
\begin{flalign}
\mathbbm{t}(\boldsymbol{x},\boldsymbol{k})=\int_{\boldsymbol{q}}c_{\boldsymbol{k}-\frac{\boldsymbol{q}}{2}}^{\dagger}c_{\boldsymbol{k}+\frac{\boldsymbol{q}}{2}}e^{\mathtt{i}\boldsymbol{q}\cdot\boldsymbol{x}}, \label{eq:_U(1)_PH_generator}
\end{flalign}
where $c$ is the gauge-invariant election operator. It generates an infinite-dimensional Lie algebra 
\begin{flalign}
[\mathbbm{t}(\xi),\mathbbm{t}(\eta)]=-2\mathtt{i}\sin\left(\frac{\partial_{\xi}\curlywedge\partial_{\eta}}{2}\right)\delta^{2d}(\xi-\eta)\mathbbm{t}(\xi), \label{eq:_U(1)_PH_algebra}
\end{flalign}
where $\xi=(\boldsymbol{x},\boldsymbol{k})$ is a coordinate in the $2d$-dimensional phase space, and the antisymmetric
product $\curlywedge$ is defined by $\xi\curlywedge\eta=\boldsymbol{\xi}^{x}\cdot\boldsymbol{\eta}^{p}-\boldsymbol{\xi}^{p}\cdot\boldsymbol{\eta}^{x}$. When $d=1$, this algebra is commonly known as the $W_{\infty}$ algebra. In higher dimensions, we refer to it as the particle-hole algebra. In the context of condensed matter systems, we assume that the phase space is compactified, with the momentum vector $\boldsymbol{k}$ living on the Brillouin zone which is a $d$-dimensional torus. The first observation is that Fermi liquids realize a condensation of the Lie-algebra generator $\mathbbm{t}$
\begin{flalign}
F_{0}(\boldsymbol{x},\boldsymbol{k})\overset{\textrm{df}}{=}\langle\textrm{FS}|\mathbbm{t}(\boldsymbol{x},\boldsymbol{k})|\textrm{FS}\rangle=\Theta(\epsilon_{F}-\epsilon_{\boldsymbol{k}}),
\end{flalign}
where $|\textrm{FS}\rangle=\prod_{\epsilon_{\boldsymbol{k}}\leq\epsilon_{F}}c_{\boldsymbol{k}}^{\dagger}|0\rangle$ is the Fermi surface (FS) ground state. This is just the distribution function $F_{0}(\boldsymbol{x},\boldsymbol{k})$ that describes the shape of FS. This is analogous to magnetic orders that realize the condensation of fermion bilinears $\langle c^{\dagger}\vec{\sigma}c\rangle$ where $\vec{\sigma}$ are Pauli matrices.

The fluctuations of the condensation $\langle\mathbbm{t}\rangle$ are systematically described by nonlinear bosonization~\cite{nlbosonization}. Given the symmetry group generated by the Lie algebra Eq.~\eqref{eq:_U(1)_PH_algebra}, one can introduce the coherent state 
\begin{flalign}
|\phi\rangle=\exp\left(\mathtt{i}\int_{\boldsymbol{x},\boldsymbol{k}}\phi(\boldsymbol{x},\boldsymbol{k})\mathbbm{t}(\boldsymbol{x},\boldsymbol{k})\right)|\textrm{FS}\rangle, \label{eq:_U(1)_coh_state}
\end{flalign}
where $\phi(\boldsymbol{x},\boldsymbol{k})$ is a bosonic variable in phase space. The distribution function dressed by fluctuations is then
\begin{flalign}
F(\boldsymbol{x},\boldsymbol{k})\overset{\textrm{df}}{=}\langle\phi|\mathbbm{t}(\boldsymbol{x},\boldsymbol{k})|\phi\rangle.
\end{flalign}
The quantization of FS fluctuations is given by the coherent-state path integral
\begin{flalign}
\mathcal{Z}=\int\mathcal{D}[\phi]e^{-\mathcal{S}[\phi]},\qquad\mathcal{S}[\phi]=\int\textrm{d}\tau\langle\phi|\partial_{\tau}+H|\phi\rangle. 
\label{eq:_U(1)_FS_action}
\end{flalign}
which can be unpacked order by order in terms of the boson $\phi$. In practical calculations~\cite{nlbosonization}, it is useful to consider a truncation of the algebra Eq.~\eqref{eq:_U(1)_PH_algebra} using the separation of energy scales $\partial_{x}\partial_{k}\sim q/k_{F}\ll1$, where $q$ is the low-energy relative momentum of particle-hole pairs. Another simplification comes from the redundancy in $\phi(\boldsymbol{x},\boldsymbol{k})$ due to the group generated by Eq.~\eqref{eq:_U(1)_PH_algebra} being partially broken by the FS ground state. The FS fluctuations are sufficiently described by $\phi(\boldsymbol{x},\boldsymbol{k}_{F})$ where $\boldsymbol{k}_{F}$ labels points on the FS manifold~\cite{nlbosonization}. The leading-order Gaussian part reproduces the well-known result based on patch assumptions
\begin{flalign}
\mathcal{S}=\frac{1}{4\pi}\int_{\boldsymbol{k}_{F}\in\textrm{FS}}\int_{\tau,\boldsymbol{x}}(\hat{\boldsymbol{v}}_{F}\cdot\partial_{\boldsymbol{x}}\phi)(\mathtt{i}\partial_{\tau}\phi+\boldsymbol{v}_{F}\cdot\partial_{\boldsymbol{x}}\phi), \label{eq:_cLL_FS}
\end{flalign}
where $\boldsymbol{v}_{F}$ is the fermi velocity and $\hat{\boldsymbol{v}}_{F}$ denotes its direction. In other words, one has a chiral Luttinger liquid on each patch of the FS in the direction $\hat{\boldsymbol{v}}_{F}$. In the full action Eq.~\eqref{eq:_U(1)_FS_action}, different patches are allowed to talk to each other. Namely, the third-order term in $\phi$ contains the gradient $\partial_{\boldsymbol{k}_{F}}\phi$ along the FS, and therefore couples the nearest neighbor patches. For a more detailed analysis of the higher-order terms in the calculation, interested readers may consult Ref.~\cite{nlbosonization}. It is worth noting that the procedure we have outlined here has parallels with the derivation of the non-linear sigma model for magnetic orders using spin coherent-state path integral methods. (see e.g.~\cite{fradkinbook} for a textbook treatment).

\end{widetext}

\bibliography{BF_CFS}

\end{document}